\documentclass[submission,copyright,creativecommons]{eptcs}
\providecommand{\event}{PLACES 2023} 

\usepackage{iftex}

\ifpdf
  \usepackage{underscore}         
  \usepackage[T1]{fontenc}        
\else
  \usepackage{breakurl}           
\fi

\usepackage{amsmath}
\usepackage{amsfonts}
\usepackage{amssymb}
\usepackage{amsthm}
\usepackage{booktabs}
\usepackage{cite}
\usepackage{graphicx}
\usepackage{stmaryrd}
\usepackage{subcaption}
\usepackage{tikz}
\usepackage{xfrac}

\usepackage{subcaption}
\usepackage{accents}

\usepackage{wrapfig}

\DeclareMathAlphabet{\mathpzc}{OT1}{pzc}{m}{it}
\DeclareMathAlphabet{\mathcal}{OMS}{cmsy}{m}{n}

\newcommand{\runiv}{\mathbb{R}}
\newcommand{\xuniv}{\mathbb{X}}
\newcommand{\vuniv}{\mathbb{V}}
\newcommand{\euniv}{\mathbb{E}}

\newcommand{\cals}{\mathcal{S}}

\newcommand{\rolex}[1]{\textsf{\upshape#1}}
\newcommand{\datax}[1]{{\renewcommand{\atxx}[2]{##1\ensuremath{\mathrm{.}}##2}\texttt{\upshape\fontdimen2\font=.25em#1}}}

\newcommand{\atxx}[2]{#1.#2}
\newcommand{\atrdxx}[2]{\atxx{\rolex{#1}}{\datax{#2}}}
\newcommand{\stores}{\mspace{2mu}\storesop\mspace{2mu}}
\newcommand{\storesop}{{:=}}
\newcommand{\unbuf}{\mspace{2mu}\unbufop\mspace{2mu}}
\newcommand{\unbufop}{{\rightarrowtriangle}}
\newcommand{\chanxx}[2]{{#1#2}}

\newcommand{\send}{\mspace{0mu}\smash{\sendop}\mspace{0mu}}
\newcommand{\sendop}{{\text{\upshape\ooalign{\phantom{?}\cr\hfil!\hfil\cr}}}}
\newcommand{\recv}{\mspace{0mu}\smash{\recvop}\mspace{0mu}}

\newcommand{\recvop}{{\text{\upshape?}}}
\newcommand{\barrxx}[3][i]{\ifthenelse{\equal{#3}{}}{#1^{\smash{#2}}}{#1{}_{\smash{#3}}^{\smash{#2}}}}

\newcommand{\testxx}[2]{#1.#2}

\newcommand{\subjx}[1]{\subjfun(#1)}
\newcommand{\subjfun}{\mathsf{subj}}

\newcommand{\puniv}{\mathbb{P}}
\newcommand{\suniv}{\mathbb{S}}
\newcommand{\cuniv}{\mathbb{C}}

\newcommand{\objfun}{\mathsf{obj}}
\newcommand{\objx}[1]{\objfun(#1)}

\newcommand{\readxx}[3][r]{#2\llbracket#3\rrbracket_{#1}}
\newcommand{\writexxx}[4][r]{#2[#3 \mapsto #4]_{#1}}

\newcommand\ubar[1]{\underaccent{\bar}{#1}}

\newcommand{\reduxx}[3][]{\mathrel{\smash{\mathrlap{\phantom{\xrightarrow[#3]{#2}}}\smash{\xrightarrow[#3]{#2}}}}}
\newcommand{\reduxxnosmash}[3][]{\mathrel{{\xrightarrow[#3]{#2}}\mathrlap{\phantom{\rightarrow}}^{#1}}}

\newcommand{\acq}{\mspace{4mu}\mathbf{acq}\mspace{4mu}}
\newcommand{\rel}{\mspace{4mu}\mathbf{rel}\mspace{4mu}}

\newcommand{\calp}{\mathcal{P}}
\newcommand{\cald}{\mathcal{D}}
\newcommand{\calc}{\mathcal{C}}

\theoremstyle{definition}
\newtheorem{definition}{Definition}

\newcommand{\auniv}{\mathbb{A}}

\newcommand{\one}{\mathbf{1}}
\newcommand{\bin}{\mspace{4mu}\binop\mspace{4mu}}
\newcommand{\binop}{{\circ}}

\newcommand{\alt}{\mspace{4mu}\altop\mspace{4mu}}
\newcommand{\altop}{{+}}
\newcommand{\async}{\mspace{4mu}\asyncop\mspace{4mu}}
\newcommand{\asyncop}{{;}}
\newcommand{\mer}{\mspace{4mu}\merop\mspace{4mu}}
\newcommand{\merop}{{\parallel}}
\newcommand{\ifxxx}[4][R]{\ifthenelse{\equal{#1}{}}{}{#1.}{\mathbf{if}}\mspace{4mu}#2\mspace{4mu}#3\mspace{4mu}#4}
\newcommand{\whilexxx}[4][R]{\ifthenelse{\equal{#1}{}}{}{#1.}{\mathbf{while}}\mspace{4mu}#2\mspace{4mu}\ifthenelse{\equal{\detokenize{#3}}{}}{}{\annotx{#3}\mspace{4mu}}#4}
\newcommand{\uifxxxxxx}[7][]{{\mathbf{if}}\mspace{4mu}#2\ifthenelse{\equal{#3}{}}{}{|_{#3}}\mspace{4mu}#4|_{#5}\mspace{4mu}#6|_{#7}}
\newcommand{\uwhilexxxx}[5][]{{\mathbf{while}}\mspace{4mu}#2\ifthenelse{\equal{#3}{}}{}{|_{#3}}\mspace{4mu}\ifthenelse{\equal{\detokenize{#4}}{}}{}{\annotx{#4}\mspace{4mu}}#5|_{\emptyset}}
\newcommand{\annotx}[2][]{\smash{\color{teal}#1\{#2\}}}

\newcommand{\proj}[1][]{\mathbin{\projop_{#1}}}
\newcommand{\projop}{{\upharpoonright}}

\newcommand{\wfxop}[1][]{{\ifthenelse{\equal{#1}{}}{\checkmark}{\checkmark_{\!\!\smash{#1}}}}}

\newcommand{\redux}[1]{\mathrel{\smash{\xrightarrow{#1}}\mathrlap{\phantom{\rightarrow}}}}
\newcommand{\reduxnosmash}[1]{\mathrel{{\xrightarrow{#1}}\mathrlap{\phantom{\rightarrow}}}}

\usepackage{amsmath}
\usepackage{amsfonts}
\usepackage{amssymb}

\usepackage{amsthm}
\usepackage{environ}
\usepackage{hyperref}
\usepackage{ifthen}
\usepackage{mathtools}
\usepackage{stmaryrd}
\usepackage{twoopt}
\usepackage{upgreek}

\newtheorem{proposition}{Proposition}

\newcommand{\dfracfrac}[2]{\text{%
  \rlap{\ensuremath{\dfrac{\phantom{#1}}{#2}}}%
  \raisebox{2pt}{\ensuremath{\dfrac{#1}{\phantom{#2}}}}%
}}

\newcommand{\tupx}[1]{(#1)}

\newcommand{\setxx}[2]{\{#1\mid#2\}}
\newcommand{\setx}[1]{\{#1\}}

\newcommand{\rulexxx}[3]{\dfrac{\begin{gathered}#2\end{gathered}}{\begin{gathered}#3\end{gathered}}\ifthenelse{\equal{#1}{}}{}{\thinspace\text{\upshape\scriptsize[\hypertarget{\detokenize{#1}}{\textsc{#1}}]}}}
\newcommand{\corulexxx}[3]{\dfracfrac{\begin{gathered}#2\end{gathered}}{\begin{gathered}#3\end{gathered}}\ifthenelse{\equal{#1}{}}{}{\thinspace\text{\upshape\scriptsize[\hypertarget{\detokenize{#1}}{\textsc{#1}}]}}}

\newcommandtwoopt{\BLOCK}[3][gathered][]{%
	\begingroup%
	#2%
	\ifthenelse{\equal{#1}{aligned}}{}{}%
	\left[\ifthenelse{\equal{#1}{}}{#3}{\ifthenelse{\equal{#1}{aligned}}{\!}{}\begin{#1}\ifthenelse{\equal{#1}{aligned}}{&}{}#3\end{#1}}\right]%
	\endgroup} 

\let\primenosmash\prime
\let\astnosmash\ast
\let\dagnosmash\dag
\let\ddagnosmash\ddag
\let\Snosmash\S
\let\Pnosmash\P
\renewcommand{\prime}{{\smash{\primenosmash}}}
\renewcommand{\ast}{{\smash{\astnosmash}}}
\renewcommand{\dag}{{\smash{\dagnosmash}}}
\renewcommand{\ddag}{{\smash{\ddagnosmash}}}
\renewcommand{\S}{{\smash{\Snosmash}}}
\renewcommand{\P}{{\smash{\Pnosmash}}}

\let\hatnosmash\hat
\renewcommand{\hat}[1]{\smash{\hatnosmash#1}}

\newcommand{\FONT}[1]{\textbf{#1}}

\newcommand{\AND}{\ \FONT{and}\ }

\newcommand{\IF}{\FONT{if:}\ }

\newcommand{\CASE}[1][]{\item\textbf{Case\ifthenelse{\equal{#1}{}}{}{ #1}:}\ }

\newcommand{\GRAMMAR}{\enspace{::=}\enspace}
\newcommand{\PIPE}{\enspace\smash{{\big|}}\enspace}

\usepackage{enumitem}
\setlistdepth{9}
\setlist[itemize,1]{label=\textbf{--}}
\setlist[itemize,2]{label=$\bullet$}
\setlist[itemize,3]{label=$\ast$}
\setlist[itemize,4]{label=$\cdot$}
\setlist[itemize,5]{label=$\cdot$}
\setlist[itemize,6]{label=$\cdot$}
\setlist[itemize,7]{label=$\cdot$}
\setlist[itemize,8]{label=$\cdot$}
\setlist[itemize,9]{label=$\cdot$}
\renewlist{itemize}{itemize}{9}

\newcommand{\SHIFT}{7pt}

\title{Choreographic Programming of Isolated Transactions}
\author{Ton Smeele
\institute{Open University of the Netherlands\\Heerlen, the Netherlands}
\and
Sung-Shik Jongmans
\institute{Open University of the Netherlands\\Heerlen, the Netherlands}
\institute{Centrum Wiskunde \& Informatica (CWI)\\Amsterdam, the Netherlands}
\email{ssj@ou.nl}
}
\def\titlerunning{Choreographic Programming of Isolated Transactions}
\def\authorrunning{T. Smeele \& S. Jongmans}
\begin{document}
\maketitle

\begin{abstract}
	Implementing distributed systems is hard; choreographic programming aims to
	make it easier. In this paper, we present the design of a new choreographic
	programming language that supports isolated transactions among overlapping sets
	of processes. The first idea is to track for every variable which processes are
	permitted to use it. The second idea is to use model checking to prove isolation.
\end{abstract}

\section{Introduction}
\label{sect:intro}

\subsection{Background: Choreographic Programming}
\label{sect:intro:backg}

\begin{wrapfigure}{r}{\linewidth*1/3}
	\vspace{-1\baselineskip}
	\centering
	\begin{tikzpicture}[x=.5cm,y=-1cm, font=\footnotesize]
		\node [inner sep=0mm] (G) at (0,0) {$G$};
		\node [inner sep=0mm] (L1) at (-1.5,1) {$L_1$};
		\node [inner sep=0mm] (L2) at (-.5,1) {$L_2$};
		\node [inner sep=0mm] (Ldots) at (.5,1) {$\cdots$};
		\node [inner sep=0mm] (Ln) at (1.5,1) {$L_n$};
		
		\draw [-stealth, rounded corners] ([yshift=-1mm]G.south) to ([yshift=1mm]L1.north);
		\draw [-stealth, rounded corners] ([yshift=-1mm]G.south) to ([yshift=1mm]L2.north);
		\draw [-stealth, rounded corners] ([yshift=-1mm]G.south) to ([yshift=1mm]Ln.north);
		
		\node [inner sep=0mm, anchor=east, minimum width=2.25cm] at (-2.25,0) {\strut {global program:}};
		\node [inner sep=0mm, anchor=east, minimum width=2.25cm] at (-2.25,.5) {\strut {projection:}};
		\node [inner sep=0mm, anchor=east, minimum width=2.25cm] at (-2.25,1) {\strut {local programs:}};
	\end{tikzpicture}
	
	\caption{Method}
	\label{fig:method}
	
	\vspace{-\baselineskip}
\end{wrapfigure}

Implementing distributed systems is hard; \textit{choreographic programming}
aims to make it easier
\cite{DBLP:journals/toplas/CarboneHY12,DBLP:conf/popl/CarboneM13,M13:phd}.
\autoref{fig:method} shows the idea.

Initially, a distributed system is written as a \textit{global program} $G$
(``the choreography''). It implements the behaviour of all processes
collectively, in a sequential programming style (easy to write, but hard to run
as a distributed system). For instance, the following global program implements
a distributed system in which, first, a data object is communicated from Alice
to Bob, and second, its hash.

\medbreak\noindent\hfill$
	G_{\rolex{a}\rolex{b}} = (\atxx{\rolex{a}}{\datax{"foo"}} \unbuf \atxx{\rolex{b}}{\datax{x}}) \async (\atxx{\rolex{a}}{\datax{hash}} \stores \datax{md5("foo")}) \async (\atxx{\rolex{a}}{\datax{hash}} \unbuf \atxx{\rolex{b}}{\datax{y}})
$\hfill\strut\medbreak

\noindent Here, $\atxx{p}{e} \unbuf \atxx{q}{y}$ and $\atxx{q}{y} \stores e$
express inter-process \textit{communication} and intra-process
\textit{computation}. Communication $\atxx{p}{e} \unbuf \atxx{q}{y}$ implements
the output of the value of expression $e$ at process $p$ and the corresponding
input into variable $y$ at process $q$; the transport is asynchronous, reliable,
and FIFO. Computation $\atxx{q}{y} \stores e$ implements the storage of the
value of expression $e$ in variable $y$ at process $q$.

Subsequently, the distributed system is run as a family of \textit{local
programs} $L_1, \ldots, L_n$, automatically extracted from the global program
through \textit{projection}. The local programs implement the behaviour of each
process individually, in a parallel programming style (easy to run as a
distributed system, but hard to write). For instance, the following local
programs implement Alice and Bob:

\medbreak\noindent\hfill$ L_{\rolex{a}} = (\chanxx{\rolex{a}}{\rolex{b}} \send
{\datax{"foo"}}) \async (\atxx{\rolex{a}}{\datax{hash}} \stores
\datax{md5("foo")}) \async (\chanxx{\rolex{a}}{\rolex{b}} \send {\datax{hash}})
\qquad L_{\rolex{b}} = (\chanxx{\rolex{a}}{\rolex{b}} \recv {\datax{x}}) \async
(\chanxx{\rolex{a}}{\rolex{b}} \recv {\datax{y}}) $\hfill\strut\medbreak

\noindent Here, \textit{send} $\chanxx{p}{q} \send e$ and \textit{receive}
$\chanxx{p}{q} \recv y$ implement an output and an input through the channel
from $p$ to $q$.

The keystone assurance of choreographic programming is \textit{operational
equivalence}: methodically, a global program and its family of local programs
are assured to have the same behaviour. To prove properties of families of local
programs, operational equivalence allows us to prove them of global programs
instead. This is typically simpler. A premier example of such a property is
\textit{absence of deadlocks}.

Choreographic programming originated with Carbone et al.
\cite{DBLP:conf/esop/CarboneHY07,DBLP:journals/toplas/CarboneHY12} (using
{binary session types} \cite{DBLP:conf/esop/HondaVK98}) and with Carbone and
Montesi \cite{DBLP:conf/popl/CarboneM13,M13:phd} (using {multiparty session
types} \cite{DBLP:conf/popl/HondaYC08}); substantial progress has been made since. Montesi and Yoshida developed a theory of compositional
choreographic programming that supports open distributed systems
\cite{DBLP:conf/concur/MontesiY13}; Carbone et al. studied connections between
choreographic programming and linear logic
\cite{DBLP:journals/dc/CarboneMS18,DBLP:conf/lopstr/CarboneCMM18}; Dalla Preda et al. combined choreographic programming with dynamic adaptation \cite{DBLP:conf/sle/PredaGLMG14,DBLP:conf/coordination/PredaGGLM15,DBLP:journals/corr/PredaGGLM16}; Cruz-Filipe and Montesi developed a minimal Turing-complete language of global programs \cite{DBLP:journals/tcs/Cruz-FilipeM20}; Cruz-Filipe et al. and Kj\ae r et al. presented techniques to extract global programs from families of local programs \cite{DBLP:conf/fossacs/Cruz-FilipeLM17,DBLP:conf/lopstr/KjaerCM22}; Giallorenzo et al. studied a correspondence between choreographic programming and multitier languages \cite{DBLP:conf/ecoop/GiallorenzoMPRS21}; Jongmans and Van den Bos combined choreographic programming with deductive verification \cite{DBLP:conf/esop/JongmansB22}; Hirsch and Garg and Cruz-Filipe et al. developed functional choreographic programming languages \cite{DBLP:journals/pacmpl/HirschG22,DBLP:conf/ictac/Cruz-FilipeGLMP22}. Other work includes results on case studies \cite{DBLP:conf/forte/Cruz-FilipeM16}, procedural abstractions \cite{DBLP:conf/forte/Cruz-FilipeM17}, asynchronous communication \cite{DBLP:conf/sac/Cruz-FilipeM17}, polyadic communication \cite{DBLP:conf/sac/Cruz-FilipeMP18,DBLP:conf/forte/HildebrandtSLDC19}, implementability \cite{DBLP:conf/forte/GiallorenzoMG18}, and formalisation\slash mechanisation in Coq \cite{DBLP:conf/ictac/Cruz-FilipeMP21,DBLP:conf/itp/Cruz-FilipeMP21,DBLP:journals/pacmpl/HirschG22}. These theoretical developments are supported in practice by several tools \cite{DBLP:conf/popl/CarboneM13,DBLP:conf/sle/PredaGLMG14,DBLP:journals/corr/PredaGGLM16,DBLP:conf/ecoop/GiallorenzoMPRS21,fm2023}.


\subsection{Open Problem: Isolated Transactions}
\label{sect:intro:problem}

Suppose we need to implement a distributed system that fulfils the following
requirements:
\begin{enumerate}\setlength\itemsep{0pt}
	\item A data object and its hash are communicated from both Alice and Carol, in
	parallel, to Bob.
	
	\item \underline{Either} Alice's data object and its hash are eventually stored
	at Bob, \underline{or} Carol's (but no mixture).
\end{enumerate}
Requirement 1 can readily be fulfilled in a choreographic programming language
with parallel composition (free interleaving), as demonstrated in the following
global program:

\medbreak\noindent\hfill$
	G_{\rolex{a}\rolex{c}\rolex{b}}^\text{v1} = G_{\rolex{a}\rolex{b}}^\text{\autoref{sect:intro:backg}} \mer G_{\rolex{c}\rolex{b}}
\qquad
	G_{\rolex{c}\rolex{b}} = (\atxx{\rolex{c}}{\datax{"bar"}} \unbuf \atxx{\rolex{b}}{\datax{x}}) \async (\atxx{\rolex{c}}{\datax{hash}} \stores \datax{md5("bar")}) \async (\atxx{\rolex{c}}{\datax{hash}} \unbuf \atxx{\rolex{b}}{\datax{y}})
$\hfill\strut\medbreak

\noindent In contrast, requirement 2 cannot be fulfilled in any choreographic
programming language that we know of (i.e., none of the choreographic
programming languages cited in \autoref{sect:intro:backg} seem to be capable of
it). What is needed, is a mechanism to run $G_{\rolex{a}\rolex{b}}$ and
$G_{\rolex{c}\rolex{b}}$ as isolated \textit{transactions}.

One possibility is to enrich the language with the standard non-deterministic
choice operator $+$. In that case, the system can be implemented as
$(G_{\rolex{a}\rolex{b}} \async
G_{\rolex{c}\rolex{b}}) \alt (G_{\rolex{c}\rolex{b}} \async
G_{\rolex{a}\rolex{b}})$. However, such an
approach, in which parallel compositions are explicitly expanded into choices,
generally leads to exponentially sized global programs (in the number of
transactions), while obscuring the intention of the system. For instance, if
Dave were added as a third client of Bob, we need to write the following global
program:

\medbreak\noindent\hfill$
	G_{\rolex{a}\rolex{b}\rolex{c}\rolex{d}}^\text{v1} = \begin{aligned}[t]
	&	
		(G_{\rolex{a}\rolex{b}} \async ((G_{\rolex{c}\rolex{b}} \async G_{\rolex{d}\rolex{b}}) \alt (G_{\rolex{d}\rolex{b}} \async G_{\rolex{c}\rolex{b}}))) \alt
	\\&
		(G_{\rolex{c}\rolex{b}} \async ((G_{\rolex{a}\rolex{b}} \async G_{\rolex{d}\rolex{b}}) \alt (G_{\rolex{d}\rolex{b}} \async G_{\rolex{a}\rolex{b}}))) \alt
	\\&
		(G_{\rolex{d}\rolex{b}} \async ((G_{\rolex{a}\rolex{b}} \async G_{\rolex{c}\rolex{b}}) \alt (G_{\rolex{c}\rolex{b}} \async G_{\rolex{a}\rolex{b}})))
	\end{aligned}
\qquad
	G_{\rolex{d}\rolex{b}} = \begin{aligned}[t]
	&
		(\atxx{\rolex{d}}{\datax{"baz"}} \unbuf \atxx{\rolex{b}}{\datax{x}}) \async
	\\&
		(\atxx{\rolex{d}}{\datax{hash}} \stores \datax{md5("baz")}) \async
	\\&	
		(\atxx{\rolex{d}}{\datax{hash}} \unbuf \atxx{\rolex{b}}{\datax{y}})
	\end{aligned}
$\hfill\strut\medbreak

\noindent Moreover, if we want to allow \textit{independent segments} of
transactions, which use disjoint sets of variables, to overlap to improve
performance (i.e., their interleaved execution would not break isolation), then
programmability is further complicated with the non-deterministic choice
approach.

To avoid these issues, we propose a more fine-grained approach in this paper
that supports eventual consistency while allowing for interleaved execution of
isolated transactions. Instead of manually implementing isolated transactions by
enumerating admissible sequences of communications, in our approach, isolation
emerges out of explicit programming language support.

\subsection{Contributions of This Paper}
\label{sect:intro:contrib}

We present the design of a new choreographic programming language that supports
isolated transactions.

The first idea is to track for each variable which processes are permitted to
use it. Initially, each process is permitted to use each variable. Subsequently,
process $p$ can \textit{acquire} exclusive permission to use variable $y$ of
process $q$. When granted, each usage of $y$ by not-$p$ is blocked until $p$
\textit{releases} its exclusive permission. Management of usage permissions is
transparant to the programmer; it is a feature of the programming language. The
following global programs demonstrate the syntax and fulfil requirement 2 in
\autoref{sect:intro:problem}:

\medbreak\noindent\hfill$\begin{alignedat}{2}
&
	G_{\rolex{a}\rolex{c}\rolex{b}}^\text{v2} &&= ((\rolex{a} \acq \atxx{\rolex{b}}{\datax{x}}) \async G_{\rolex{a}\rolex{b}}^\text{\autoref{sect:intro:backg}} \async (\rolex{a} \rel \atxx{\rolex{b}}{\datax{x}})) \mer ((\rolex{c} \acq \atxx{\rolex{b}}{\datax{x}}) \async G_{\rolex{c}\rolex{b}}^\text{\autoref{sect:intro:problem}} \async (\rolex{c} \rel \atxx{\rolex{b}}{\datax{x}}))
\\&
	G_{\rolex{a}\rolex{c}\rolex{d}\rolex{b}}^\text{v2} &&= G_{\rolex{a}\rolex{c}\rolex{b}}^\text{v2} \mer ((\rolex{d} \acq \atxx{\rolex{b}}{\datax{x}}) \async G_{\rolex{d}\rolex{b}}^\text{\autoref{sect:intro:problem}} \async (\rolex{d} \rel \atxx{\rolex{b}}{\datax{x}}))
\end{alignedat}$\hfill\strut\medbreak

\noindent We note that $G_{\rolex{a}\rolex{c}\rolex{d}\rolex{b}}^\text{v2}$ is \textit{compositionally constructed} out of
$G_{\rolex{a}\rolex{c}\rolex{b}}^\text{v2}$, without the need to refer to sub-programs $G_{\rolex{a}\rolex{b}}$
and $G_{\rolex{c}\rolex{b}}$; this is not
possible when parallel compositions are explicitly expanded into choices.

Thus, the idea of tracking usage permissions---and blocking those usages that
are forbidden---enables the programmer to write more compact global programs,
intended to better preserve the intention of the system. However:

\begin{itemize}
	\item This feature does not guarantee isolation by itself; it is just a means
	to achieve it. In other words, a separate mechanism is still needed to check
	isolation and guarantee it is preserved by projection.

	\item ``Blocking those usages that are forbidden'' also has an
	adverse side-effect: processes that compete to acquire permission to use the
	same variables can deadlock. For instance, the following global program
	implements a system in which Alice tries to acquire permission to use variables
	$\datax{x}$ and $\datax{y}$ of Bob, while Carol tries to acquire permission to
	use the same variables, but in reverse:

	\medbreak\noindent\hfill$
		(\rolex{a} \acq \atxx{\rolex{b}}{\datax{x}} \async \rolex{a} \acq \atxx{\rolex{b}}{\datax{y}} \async {\cdots}) \mer (\rolex{c} \acq \atxx{\rolex{b}}{\datax{y}} \async \rolex{c} \acq \atxx{\rolex{b}}{\datax{x}} \async {\cdots})
	$\hfill\strut\medbreak

	\noindent A deadlock arises when Alice acquires permission to use $\datax{x}$,
	while Carol acquires permission to use $\datax{y}$, so neither one of them can
	acquire permission to use a second variable.\footnote{We
	note that this is a different source of deadlock than the \textit{communication
	deadlocks} that choreographic programming traditionally avoids (i.e., waiting
	for a message that is never sent).}
\end{itemize}

\noindent To address these points, the second idea of this paper is to specify
properties, such as isolation and absence of deadlocks, in temporal logic and
use model checking to prove that they are satisfied. We believe this combination
with choreographic programming is new.

\section{The Design}
\label{sect:lang}

\begin{wrapfigure}{r}{.5\linewidth}
	\vspace{-1\baselineskip}
	\centering
	\begin{tikzpicture}[x=2.5cm, y=-.5cm, minimum height=.5cm-1mm, minimum width=7.5cm-1mm, inner sep=0pt, font=\footnotesize\strut]
		\node [draw] at (1,0) {systems (\autoref{sect:lang:sys})};
		\node [draw, minimum width=2.5cm-1mm] at (0,1) {programs (\autoref{sect:lang:progs})};
		\node [draw, minimum width=2.5cm-1mm] at (1,1) {stores (\autoref{sect:lang:stores})};
		\node [draw, minimum width=2.5cm-1mm] at (2,1) {channels (\autoref{sect:lang:chans})};
		\node [draw] at (1,2) {actions (\autoref{sect:lang:acts})};
		\node [draw, minimum width=7.5cm*1/2-1mm] at (1/4,3) {names (\autoref{sect:lang:names})};
		\node [draw, minimum width=7.5cm*1/2-1mm] at (2-1/4,3) {data (\autoref{sect:lang:exprs})};
	\end{tikzpicture}
	
	\caption{Design}
	\label{fig:design}
	
	\vspace{-\baselineskip}
\end{wrapfigure}

We define a language in which both global programs and families of local
programs can be expressed. \autoref{fig:design} shows the design. It has four
layers: every \textit{system} is defined in terms of \textit{programs} (either a
single global one, or multiple local ones), \textit{stores} (one for every
process), and \textit{channels} (one between every pair of processes); every
program, store, or channel is defined in terms of \textit{actions},
process\slash channel \textit{names}, and \textit{data}; every action is itself
defined in terms of names and data, too.

\subsection{Names and Data}
\label{sect:lang:names}
\label{sect:lang:exprs}

First, we define: the syntax of names (\autoref{defn:runiv}); the syntax of data
(\autoref{defn:euniv}). As the topic of interest is ``processes that
communicate'', instead of ``data that are communicated'', we omit most details.

\begin{definition}\label{defn:runiv}
	Let $\runiv = \setx{\rolex{a}, \rolex{b}, \rolex{c}, \ldots}$ denote the set of
	\textit{process names}, ranged over by $p, q, r$. Let $\runiv \times \runiv
	\setminus \setxx{\tupx{r, r}}{r \in R}$ denote the set of \textit{channel
	names}. \qed
\end{definition}

\begin{definition}\label{defn:euniv}
	Let $\xuniv = \setx{\datax{\_}, \datax{x}, \datax{y}, \datax{z}, \ldots}$
	denote the set of \emph{variables}, ranged over by $x, y, z$. Let $\vuniv =
	\setx{\datax{unit},\allowbreak \datax{true}, \datax{false},\allowbreak
	\datax{0},\allowbreak \datax{1},\allowbreak \datax{2},\allowbreak \ldots,
	\datax{acq}, \datax{rel}}$ denote the set of \emph{values}, ranged over by $u,
	v, w$. Let $\euniv$ denote the set of \emph{expressions}, ranged over by $E$;
	it is defined as follows:
	
	\medbreak\noindent\hfill$
		E \GRAMMAR x \PIPE u \PIPE \datax{$E_1$\,==\,$E_2$} \PIPE \datax{\raisebox{.5ex}{\texttildelow}$E$} \PIPE \datax{$E_1$\,\&\&\,$E_2$} \PIPE \datax{$E_1$\,+\,$E_2$} \PIPE {\cdots}
	$\hfill\llap{\strut\qed}
\end{definition}

\noindent Symbol $\datax{\_}$ is a special variable that loses all data written
to it, similar to \datax{/dev/null} in Unix. Symbols $\datax{acq}$ and
$\datax{rel}$ are special values to control usage permissions of variables
(\autoref{sect:lang:stores}).

\subsection{Actions}
\label{sect:lang:acts}

Next, we define: the syntax of actions that processes can execute
(\autoref{defn:auniv}); functions to retrieve the ``subject'' and the ``object''
of an action (\autoref{defn:subj}). The subject is the process that executes an
action; the object is the channel through which an action is executed, if any.

\begin{definition}\label{defn:auniv}
	Let $\auniv$ denote the set of \emph{actions}, ranged over by $\alpha$; it is
	defined as follows:
	
	\medbreak\noindent\hfill$
		\alpha \GRAMMAR \testxx{p}{E} \PIPE \atxx{q}{y} \stores E \PIPE \chanxx{p}{q} \send E \PIPE \chanxx{p}{q} \recv E \PIPE \uptau
	$\hfill\llap{\strut\qed}
\end{definition}

\noindent Action $\testxx{p}{E}$ implements a \textit{test} of expression $E$ at
process $p$. Action $\atxx{q}{y} \stores E$ implements an \textit{assignment} of
the value of expression $E$ to variable $y$ at process $q$. Actions
$\chanxx{p}{q} \send E$ and $\chanxx{p}{q} \recv E$ implement an
\textit{asynchronous} \textit{send} and \textit{receive} of the value of
expression $E$ from process $p$ to process $q$. Action $\uptau$ implements
\textit{idling}.

\begin{definition}\label{defn:subj}
	Let $\subjx{\alpha}$ and $\objx{\alpha}$ denote the \textit{subject} and the
	\textit{object} of $\alpha$; they are defined as follows:
	
	\medbreak\noindent\hfill$
		\begin{alignedat}[b]{2}
		&
			\subjx{\testxx{p}{E}} &&= p
		\\&
			\subjx{\chanxx{p}{q} \send E} &&= p
		\end{alignedat}
	\qquad
		\begin{alignedat}[b]{2}
		&
			\subjx{\atxx{q}{y} \stores E} &&= q
		\\&
			\subjx{\chanxx{p}{q} \recv y} &&= q
		\end{alignedat}
	\qquad
		\begin{alignedat}[b]{2}
		&
			\objx{\chanxx{p}{q} \send E} &&= \chanxx{p}{q}
		\\&
			\objx{\chanxx{p}{q} \recv E} &&= \chanxx{p}{q}
		\end{alignedat}
	$\hfill\llap{\strut\qed}
\end{definition}

\subsection{Programs}
\label{sect:lang:progs}

Next, we define: the syntax of programs (\autoref{defn:puniv}); a function to
extract local programs from a global program (\autoref{defn:proj}); the
operational semantics of programs (\autoref{defn:redu:p}).

\begin{definition}\label{defn:puniv}
	Let $\puniv$ denote the set of \emph{programs}, ranged over by $P, G, L$; it is
	defined as follows:
	
	\medbreak\noindent\hfill$
		P \GRAMMAR \one \PIPE \alpha \PIPE P_1 \alt P_2 \PIPE P_1 \mer P_2 \PIPE P_1 \async P_2
	$\hfill\llap{\strut\qed}
\end{definition}

\noindent Program $\one$ implements an \textit{empty execution}. Program $P_1
\alt P_2$ implements a \textit{choice} between $P_1$ and $P_2$. Program $P_1
\mer P_2$ implements an \textit{interleaving} of $P_1$ and $P_2$. Program $P_1
\async P_2$ implements a \textit{sequence} of $P_1$ and $P_2$. Furthermore, we
use the following shorthand notation:

\medbreak\noindent\hfill$
	\begin{aligned}
		\atxx{p}{E} \unbuf \atxx{q}{y} &\enspace\text{instead of}\enspace \chanxx{p}{q} \send E \async \chanxx{p}{q} \recv y
	\\
		p \acq \atxx{q}{y} &\enspace\text{instead of}\enspace (\atxx{p}{\datax{acq}} \unbuf \atxx{q}{y}) \async (\atxx{q}{\datax{unit}} \unbuf \atxx{p}{\_})
	\\
		p \acq \atxx{q}{[y_1, \ldots, y_n]} &\enspace\text{instead of}\enspace p \acq \atxx{q}{y_1} \async {\cdots} \async p \acq \atxx{q}{y_n}
	\\
		p \rel \atxx{q}{y} &\enspace\text{instead of}\enspace \atxx{p}{\datax{rel}} \unbuf \atxx{q}{y}
	\\
		p \rel \atxx{q}{[y_1, \ldots, y_n]} &\enspace\text{instead of}\enspace p \rel \atxx{q}{y_1} \async {\cdots} \async p \rel \atxx{q}{y_n}
	\\
		\ifxxx[]{\testxx{p}{e}}{P_1}{P_2} &\enspace\text{instead of}\enspace
		(\testxx{p}{E} \async P_1) \alt (\testxx{p}{\datax{\raisebox{.5ex}{\texttildelow}$E$}} \async P_2)
	\end{aligned}
$\hfill\strut\medbreak

A program is \textit{global} if at least two subjects occur in it; it is
\textit{local} if it at most one subject occurs in it. A local program for
process $r$ can be extracted from global program $G$ through projection. The
idea is to replace every action in $G$ of which $r$ \underline{is not} the
subject with $\uptau$.

\begin{definition}\label{defn:proj}
	Let $P \proj r$ denote the \textit{projection} of $P$ onto $r$; it is induced
	by the following equations:
	
	\medbreak\noindent\hfill$
		\begin{alignedat}[b]{2}
		&
			\alpha \proj \subjfun(\alpha) &&= \alpha
		\\&
			\alpha \proj r &&= \uptau \quad\IF r \neq \subjfun(\alpha)
		\end{alignedat}
	\qquad
		\begin{aligned}[b]
			\one \proj r &= \one
		\\
			P_1 \bin P_2 \proj r &= (P_1 \proj r) \bin (P_2 \proj r) \quad\IF \binop \in \setx{\altop, \merop, \asyncop}
		\end{aligned}
	$\hfill\llap{\strut\qed}
\end{definition}

We define the operational semantics of programs through a labelled reduction
relation.

\begin{figure}[t]
	\begin{minipage}{\linewidth}\centering
		\tikz{\node[inner sep=1.5mm, minimum width=\textwidth, fill=black!5, rounded corners]{$\begin{gathered}
			\dfrac{
			}{
				\alpha \reduxnosmash{\alpha} \one
			}
		\qquad
			\dfrac{
				P_1 \reduxnosmash{\alpha} P_1'
			}{
				P_1 \alt P_2 \reduxnosmash{\alpha} P_1'
			}
		\qquad
			\dfrac{
				P_2 \reduxnosmash{\alpha} P_2'
			}{
				P_1 \alt P_2 \reduxnosmash{\alpha} P_2'
			}
		\qquad
			\dfrac{
				P_1 \reduxnosmash{\alpha} P_1'
			}{
				P_1 \mer P_2 \reduxnosmash{\alpha} P_1' \mer P_2
			}
		\qquad
			\dfrac{
				P_2 \reduxnosmash{\alpha} P_2'
			}{
				P_1 \mer P_2 \reduxnosmash{\alpha} P_1 \mer P_2'
			}
		\\[\SHIFT]
			\dfrac{
				P_1 \reduxnosmash{\alpha} P_1'
			}{
				P_1 \async P_2 \reduxnosmash{\alpha} P_1' \async P_2
			}
		\qquad
			\dfrac{
				\subjx{\alpha} \notin \setxx{\subjx{\hat\alpha}}{{P_1 \rightarrow {\cdots} \rightarrow \redux{\hat\alpha}}}
			\qquad
				P_2 \reduxnosmash{\alpha} P_2'
			}{
				P_1 \async P_2 \reduxnosmash{\alpha} P_2'
			}
		\end{gathered}$}}
		
		\subcaption{Programs. Let $\rightarrow {\cdots} \rightarrow$ denote a sequence of $0$-or-more reductions.}
		\label{fig:semantics:prog}
	\end{minipage}
	
	\bigbreak
	
	\begin{minipage}{\linewidth}\centering
		\tikz{\node[inner sep=1.5mm, minimum width=\textwidth, fill=black!5, rounded corners]{$\begin{gathered}
			\dfrac{
				\readxx[p]{S}{E} = \datax{true}
			}{
				S \reduxxnosmash{\testxx{p}{E}}{\testxx{p}{\datax{true}}} S
			}
		\qquad
			\dfrac{
				\readxx[q]{S}{E} = v
			}{
				S \reduxxnosmash{\atxx{q}{y} \stores E}{\atxx{q}{y} \stores v} \writexxx[q]{S}{y}{v}
			}
		\qquad
			\dfrac{
				\readxx[p]{S}{E} = u
			}{
				S \reduxxnosmash{\chanxx{p}{q} \send E}{\chanxx{p}{q} \send u} S
			}
		\qquad
			\dfrac{
			}{
				S \reduxxnosmash{\chanxx{p}{q} \recv y\mathrlap{\phantom{E}}}{\chanxx{p}{q} \recv v} \writexxx[p]{S}{y}{v}
			}
		\qquad
			\dfrac{
			}{
				S \reduxxnosmash{\uptau\mathrlap{\phantom{E}}}{\uptau} S
			}
		\end{gathered}$}}
		
		\subcaption{Stores}
		\label{fig:semantics:store}
	\end{minipage}
	
	\bigbreak
	
	\begin{minipage}{\linewidth}\centering
		\tikz{\node[inner sep=1.5mm, minimum width=\textwidth, fill=black!5, rounded corners]{$\begin{gathered}
			\dfrac{
			}{
				C \reduxxnosmash{}{\testxx{p}{v}} C
			}
		\qquad
			\dfrac{
			}{
				C \reduxxnosmash{}{\atxx{q}{y} \stores v} C
			}
		\qquad
			\dfrac{
				|\vec v| < n
			}{
				\tupx{\vec v, n} \reduxxnosmash{}{\chanxx{p}{q} \send u} \tupx{u{\cdot}\vec v, n}
			}
		\qquad
			\dfrac{
			}{
				\tupx{\vec u {\cdot} v, n} \reduxxnosmash{}{\chanxx{p}{q} \recv v} \tupx{\vec u, n}
			}
		\qquad
			\dfrac{
			}{
				C \reduxxnosmash{}{\uptau} C
			}
		\end{gathered}$}}
		
		\subcaption{Channels}
		\label{fig:semantics:chan}
	\end{minipage}
	
	\bigbreak
	
	\begin{minipage}{\linewidth}\centering
		\tikz{\node[inner sep=1.5mm, minimum width=\textwidth, fill=black!5, rounded corners]{$\begin{gathered}
			\dfrac{
				P \reduxnosmash{\alpha} P'
			}{
				\setx{P} \cup \mathcal{P} \reduxnosmash{\alpha} \setx{P'} \cup \mathcal{P}
			}
		\qquad
			\dfrac{
				S \reduxxnosmash{\alpha}{\ubar\alpha} S'
			\qquad
				\subjfun(\alpha) = r
			}{
				\setx{\subjfun(\alpha) \mapsto S} \cup \mathcal{S} \reduxxnosmash{\alpha}{\ubar\alpha} \setx{\subjfun(\alpha) \mapsto S'} \cup \mathcal{S}
			}
		\\[\SHIFT]
			\dfrac{
				C \reduxxnosmash{}{\ubar\alpha} C'
			\qquad
				\objx{\ubar\alpha} = \chanxx{p}{q}
			}{
				\setx{\chanxx{p}{q} \mapsto C} \cup \mathcal{C} \reduxxnosmash{\phantom{\alpha}}{\ubar\alpha} \setx{\chanxx{p}{q} \mapsto C'} \cup \mathcal{C}
			}
		\qquad
			\dfrac{
				\mathcal{P} \reduxxnosmash{\alpha}{} \mathcal{P'}
			\qquad
				\mathcal{S} \reduxxnosmash{\alpha}{\ubar\alpha} \mathcal{S'}
			\qquad
				\mathcal{C} \reduxxnosmash{}{\ubar\alpha} \mathcal{C'}
			}{
				\tupx{\mathcal{P}, \mathcal{S}, \mathcal{C}} \reduxxnosmash{\alpha}{\ubar \alpha} \tupx{\mathcal{P}', \mathcal{S}', \mathcal{C}'}
			}
		\end{gathered}$}}
		
		\subcaption{Systems}
		\label{fig:semantics:sys}
	\end{minipage}
	
	\caption{Operational semantics}
	\label{fig:semantics}
\end{figure}

\begin{definition}\label{defn:redu:p}
	Let $P \reduxnosmash{\alpha} P'$ denote \textit{reduction} from $P$ to $P'$
	with $\alpha$; it is defined in \autoref{fig:semantics:prog}. \qed
\end{definition}

\noindent Most rules are standard. The only special rule is the second rule for
sequencing: it allows sequences of actions to be executed \textit{out-of-order},
so long as they are executed at different processes (i.e., they are independent;
insisting on a sequential order would be unreasonable in a parallel
environment). That is, the left premise of the rule entails that the subject of
$\alpha$ does not occur in $P_1$ (cf. the operational semantics of global
multiparty session types). For instance, in $\atrdxx{a}{x} \stores \datax{5}
\async \atrdxx{b}{y} \stores \datax{6}$, the assignments at Alice and Bob
\underline{\smash{may}} be executed out-of-order. In contrast, in $\atrdxx{a}{x}
\stores \datax{5} \async \atrdxx{a}{x+1} \unbuf \atrdxx{b}{y}$, the assignment
and the communication \underline{must} be executed in-order.

\subsection{Stores}
\label{sect:lang:stores}

Next, we define: the syntax of stores (\autoref{defn:suniv}); functions to read
expressions from a store and write values to it (\autoref{defn:read}); the
operational semantics of stores (\autoref{defn:redu:s}).

\begin{definition}\label{defn:suniv}
	Let $\suniv = (\xuniv \setminus \setx{\datax{\_}}) \rightharpoonup (\vuniv
	\times 2^\runiv)$ denote the set of \textit{stores}, ranged over by $S$. \qed
\end{definition}

\noindent Storage $S(x) = \tupx{u, R}$ means that variable $x$ has value $u$,
and that the processes in $R$ are permitted to use it. Typically, $R \in
\setx{\runiv} \cup \setxx{\setx{r}}{r \in \runiv}$: either every process is
permitted to use $x$ (if $R = \runiv$), or only one process (if $R = \setx{r}$
for some $r \in \runiv$). Every process has its own store, but through
communications, other processes can use it, too.

\begin{definition}\label{defn:read}
	Let $\readxx{S}{E}$ and $\writexxx{S}{y}{v}$ denote the \textit{read} of $E$ in
	$S$ by $r$ and the \textit{write} of $v$ to $y$ in $S$ by $r$; they are defined as follows:

	\medbreak\noindent\hfill$
		\begin{alignedat}[b]{2}
		&
			\readxx{S}{x} &&= u \quad\IF S(x) = \tupx{u, R} \AND r \in R
		\\&
			\readxx{S}{u} &&= u
		\\&
			\writexxx{S}{\_}{v} &&= S
		\\&
			\writexxx{S}{y}{v} &&= \smash{\setxx{x \mapsto S(x)}{x \neq y} \cup \begin{cases}
				\setx{y \mapsto \tupx{v, R}} &\IF \datax{acq} \neq v \neq \datax{rel}
			\\
				\setx{y \mapsto \tupx{u, \setx{r}}} &\IF \datax{acq} = v \neq \datax{rel}
			\\
				\setx{y \mapsto \tupx{u, \runiv}} &\IF \datax{acq} \neq v = \datax{rel}
			\end{cases}}
		\\&
		\\&
			&&\phantom{{}={}} \IF y \neq \datax{\_} \AND S(y) = \tupx{u, R} \AND r \in R
		\end{alignedat}
	\qquad
		\begin{alignedat}[b]{2}
		&
			\readxx{S}{\datax{$E_1$\,==\,$E_2$}} &&= {\ldots}
		\\&
			\readxx{S}{\datax{\raisebox{.5ex}{\texttildelow}$E$}} &&= {\ldots}
		\\&
			\readxx{S}{\datax{$E_1$\,\&\&\,$E_2$}} &&= {\ldots}
		\\&
			\readxx{S}{\datax{$E_1$\,+\,$E_2$}} &&= {\ldots}
		\\&
			\ooalign{\phantom{$\readxx{S}{\datax{$E_1$\,\&\&\,$E_2$}}$}\cr\hfil$\smash{\vdots}$\hfil} && \ooalign{\phantom{${}={}$}\cr\hfil$\smash{\vdots}$\hfil} \ooalign{\phantom{${\ldots}$}\cr\hfil$\smash{\vdots}$\hfil}
		\\&
		\end{alignedat}
	$\hfill\llap{\strut\qed}
\end{definition}

\noindent Writes $\writexxx{S}{y}{\datax{acq}}$ and
$\writexxx{S}{y}{\datax{rel}}$ mean that process $r$ tries to acquire or release
exclusive permission to use $y$, without changing the value; it succeeds only if
$r$ already has permission (possibly non-exclusive).

The crux of the definition is that $\readxx{S}{E}$ and $\writexxx{S}{y}{v}$ are
undefined when $r$ is not permitted to use a variable that occurs in $E$ or $y$.
Such undefinedness is leveraged in the operational semantics of stores (next
definition). We note that $\readxx{S}{E}$ is also undefined when operations are
performed on sub-expressions of incompatible types. For instance,
$\readxx{S}{\datax{5\,+\,true}}$ is undefined. A type system can be used to
catch such errors statically; this is orthogonal to the aim of this paper.

We define the operational semantics of stores through a labelled reduction
relation. Every reduction has two labels: an action (written above the arrow)
and the ``ground'' version of the action (written below). In the ground version,
every expression is replaced by its value, if any.

\begin{definition}\label{defn:redu:s}
	Let $S \reduxxnosmash{\alpha}{\ubar\alpha} S'$ denote \textit{reduction} from
	$S$ to $S'$ with $\alpha$ and $\ubar\alpha$; it is defined in
	\autoref{fig:semantics:store}. \qed
\end{definition}

\noindent The first rule states that a test $\testxx{p}{E}$ is executed on a
store by reading $E$, if the value of $E$ is $\datax{true}$, and if $p$ has
enough permissions. The second rule states that an assignment $\atxx{q}{y}
\stores E$ is executed by reading $E$, and by writing the value of $E$ to $y$,
if $q$ has enough permissions. The third rule states that a send $\chanxx{p}{q}
\send E$ is executed by reading $E$, if $p$ has enough permissions. The fourth
rule states that a receive $\chanxx{p}{q} \recv y$ and its ground version
$\chanxx{p}{q} \recv v$ are executed by writing $v$ to $y$, if $p$  has
permission to use $y$ (not $q$; essentially, we treat receives as remote
assignments). If a process does not have enough permissions for a rule to be
applicable, the store cannot reduce, so the action is blocked.

\subsection{Channels}
\label{sect:lang:chans}

Next, we define: the syntax of channels (\autoref{defn:cuniv}); the operational
semantics (\autoref{defn:redu:c}). Henceforth, we write $\vec u$ for a list of
values, and we write $v{\cdot}\vec u$ and $\vec u{\cdot}v$ for prefixing and
suffixing.

\begin{definition}\label{defn:cuniv}
	Let $\cuniv = \vuniv^* \times \setx{0, 1, 2,\allowbreak \ldots, \infty}$ denote
	a set of \textit{channels}, ranged over by $C$. \qed
\end{definition}

\noindent Channel $\tupx{\vec u, n}$ means that its $n$-capacity buffer contains
the values in $\vec u$; the buffer is reliable and FIFO.

We define the operational semantics of channels through a labelled reduction
relation. As channels contain values, every reduction has one label: a ground
action (written below the arrow).

\begin{definition}\label{defn:redu:c}
	Let $C \reduxxnosmash{}{\ubar\alpha} C'$ denote \textit{reduction} from
	$C$ to $C'$ with $\ubar\alpha$; it is defined in \autoref{fig:semantics:chan}.
	\qed
\end{definition}

\noindent The first and second rule state that a test and an assignment are
executed on a channel without really using it. The third rule states that a send
is executed by enqueueing a value to the buffer, if it is not full. The fourth
rule states that a receive is executed by dequeueing a value from the buffer, if
it is not empty. Henceforth, we omit reduction labels when they do not matter.

\subsection{Systems}
\label{sect:lang:sys}

Last, we define: the syntax of systems (\autoref{defn:systems}); the operational
semantics (\autoref{defn:redu:systems}); operational equivalence (\autoref{defn:equiv})

\begin{definition}\label{defn:systems}
	Let $\pmb\puniv = 2^\puniv \setminus \setx{\emptyset}$ denote the set of
	(non-empty) \textit{sets of programs}, ranged over by $\calp$. Let $\pmb\suniv
	= \runiv \rightharpoonup \mathbb{S}$ denote the set of \textit{families of
	stores}, ranged over by $\mathcal{S}$. Let $\pmb\cuniv = \runiv \times \runiv
	\rightharpoonup \mathbb{C}$ denote the set of \textit{families of channels},
	ranged over by $\mathcal{C}$. Let $\pmb\puniv \times \pmb\suniv \times
	\pmb\cuniv$ denote the set of \textit{systems}, ranged over by $\cald$. \qed
\end{definition}

\noindent System $\tupx{\calp, \cals, \calc}$ means that the program(s) in
$\calp$, the stores in $\cals$, and the channels in $\calc$ are executed
together. It is well-formed if there exists a set of processes $R = \setx{r_1,
\ldots, r_n}$ such that the domain of $\cals$ is $R$ (every process has a
store), and the domain of $\calc$ is $R \times R$ (every pair of processes has a
channel), and:

\medbreak\noindent\hfill$
	\begin{aligned}
		\calp &\in \setxx{\setx{P}}{P \text{ is global and every subject that occurs in $P$ occurs in $R$}} \cup {}
	\\
		&{\phantom{{}\in{}}} \setxx{\setx{P_1, \ldots, P_n}}{\text{for each $1 \leq i \leq n$, $P_{r_i}$ is local and every subject that occurs in $P_{r_i}$ is $r_i$}}
	\end{aligned}
$\hfill\strut\medbreak

We define the operational semantics of systems through a labelled reduction
relation.

\begin{definition}\label{defn:redu:systems}
	Let $\tupx{\calp, \cals, \calc} \reduxx{\alpha}{\ubar\alpha}
	\tupx{\calp, \cals, \calc}'$ denote \textit{reduction} from $\tupx{\calp,
	\cals, \calc}$ to $\tupx{\calp, \cals, \calc}$ with $\alpha$ and $\ubar\alpha$;
	it is defined in \autoref{fig:semantics:sys}. \qed
\end{definition}

\noindent The first, second, and third rule lift reduction from individual
programs, stores, and channels to sets of programs, families of stores, and
families of channels. The fourth rule connects them together.

Two systems are operationally equivalent if they have the same behaviour. We
formalise ``having the same behaviour'' in terms of \textit{branching
bisimilarity} \cite{DBLP:journals/jacm/GlabbeekW96} (in contrast to trace
equivalence as usual), because: it is insensitive to idling; it preserves the
validity of formulas in many temporal logics (including LTL, CTL, CTL$^*$, and
$\mu$-calculus, subject to conditions), which we require to specify properties
of global programs. Two systems (resp. processes, stores, channels, sets of
processes, families of stores, families of channels) are branching bisimilar iff
they can repeatedly mimic each other's reductions, modulo idling.

\begin{definition}\label{defn:equiv}
	Let $\setx{{\approx_1}, {\approx_2}, \ldots}$ denote the set of
	\textit{branching bisimulations}, ranged over by $\approx$; it is defined as
	follows, coinductively:
	
	\medbreak\noindent\hfill$
		\begin{gathered}
			\dfrac{
				\begin{gathered}
					\bullet\ \text{for each $\cald_1 \reduxxnosmash[*]{\uptau}{\uptau} \cald_1^\dag \reduxxnosmash{\alpha}{\ubar\alpha} \cald_1^\ddag \reduxxnosmash[*]{\uptau}{\uptau} \cald_1'$, for some $\cald_2 \reduxxnosmash[*]{\uptau}{\uptau} \cald_2^\dag \reduxxnosmash{\alpha}{\ubar\alpha} \cald_2^\ddag \reduxxnosmash[*]{\uptau}{\uptau} \cald_2'$, $\cald_1^\dag \approx \cald_2^\dag$, $\cald_1^\ddag \approx \cald_2^\ddag$, $\cald_1' \approx \cald_2'$}
				\\
					\bullet\ \text{for each $\cald_2 \reduxxnosmash[*]{\uptau}{\uptau} \cald_2^\dag \reduxxnosmash{\alpha}{\ubar\alpha} \cald_2^\ddag \reduxxnosmash[*]{\uptau}{\uptau} \cald_2'$, for some $\cald_1 \reduxxnosmash[*]{\uptau}{\uptau} \cald_1^\dag \reduxxnosmash{\alpha}{\ubar\alpha} \cald_1^\ddag \reduxxnosmash[*]{\uptau}{\uptau} \cald_1'$, $\cald_1^\dag \approx \cald_2^\dag$, $\cald_1^\ddag \approx \cald_2^\ddag$, $\cald_1' \approx \cald_2'$}
				\end{gathered}
			}{
				\cald_1 \approx \cald_2
			}
		\end{gathered}
	$\hfill\strut\medbreak
	
	\noindent Let ${\equiv} = {\approx_1} \cup {\approx_2} \cup {\cdots}$ denote
	\textit{operational equivalence} (i.e., the largest branching bisimulation).
	\qed
\end{definition}

\noindent The following proposition states that operational equivalence of sets
of programs implies that of the systems they constitute. Specifically, if $P$ is
a global program, and if $\setx{P} \equiv \setxx{P \proj r}{\text{$r$ is a
subject of $P$}}$, then the local programs extracted from $P$ have the same
behaviour as $P$ in \textit{any} initial stores and channels. In the absence of
loops, as in this paper, checking $\calp_1 \equiv \calp_2$ is clearly decidable;
in the presence of loops, it is not. We leave decidable approximations of
$\equiv$ (e.g., well-formedness conditions on the syntax of choices, as usual)
for future work, when we extend our work with loops.

\begin{proposition}
	For all $\cals, \calc$, if $\calp_1 \equiv \calp_2$, then $\tupx{\calp_1,
	\cals, \calc} \equiv \tupx{\calp_2, \cals, \calc}$. \qed
\end{proposition}

\subsection{Properties}
\label{sect:lang:props}

To prove properties, we adopt a state-based temporal logic in the style of CTL
\cite{DBLP:journals/scp/EmersonC82}. We are primarily interested in two classes
of properties (although other classes may be specified, too): isolation of
transactions and absence of deadlock; our logic has special predicates to
formulate such properties. The need to explicitly prove absence of deadlock
arises from the fact that systems in this paper are not deadlock-free by
construction. For instance, any system that consists of the following program
can deadlock (elaboration of the last example in \autoref{sect:intro:contrib}):

\medbreak\noindent\hfill$
	G_{\rolex{a}\rolex{c}\rolex{b}}^\text{v3} = ((\rolex{a} \acq \atxx{\rolex{b}}{[\datax{x}, \datax{y}]}) \async G_{\rolex{a}\rolex{b}}^\text{\autoref{sect:intro:backg}} \async (\rolex{a} \rel \atxx{\rolex{b}}{[\datax{x}, \datax{y}]})) \mer ((\rolex{c} \acq \atxx{\rolex{b}}{[\datax{y}, \datax{x}]}) \async G_{\rolex{c}\rolex{b}}^\text{\autoref{sect:intro:problem}} \async (\rolex{c} \rel \atxx{\rolex{b}}{[\datax{x}, \datax{y}]}))
$\hfill\strut\medbreak

\noindent The problem is that Alice acquires $\datax{x}$ and $\datax{y}$ (in
that order), while Carol acquires $\datax{y}$ and $\datax{x}$ (in that order).

\begin{definition}
	Let $\mathbb{F}$ denote the set of \textit{formulas}, ranged over by $\varphi$;
	it is defined as follows:
	
	\medbreak\noindent\hfill$
		\varphi \GRAMMAR \top \PIPE \neg \varphi \PIPE \varphi_1 \wedge \varphi_2 \PIPE \mathsf{EG}(\varphi) \PIPE \mathsf{EU}(\varphi_1, \varphi_2) \PIPE \testxx{p}{E} \PIPE \mathsf{AX}_{\atxx{q}{y}}(\varphi) \PIPE \mathsf{dead}
	$\hfill\llap{\strut\qed}
\end{definition}

\noindent Formula $\top$ specifies \textit{truth}. Formulas $\neg\varphi$ and
$\varphi_1 \wedge \varphi_2$ specify \textit{negation} and \textit{conjunction}.
Formula $\mathsf{EG}(\varphi)$ specifies that, in some branch, $\varphi$ is
\textit{always} true. Formula $\mathsf{EU}(\varphi_1, \varphi_2)$ specifies
that, in some branch, $\varphi_1$ is true \textit{until} $\varphi_2$ is true.
Formula $\testxx{p}{E}$ specifies \textit{proposition} $E$ at $p$. Formula
$\mathsf{AX}_{\atxx{q}{y}}(\varphi)$ specifies that $\varphi$ is true
\textit{next} if variable $y$ at process $q$ was changed. Formula
$\mathsf{dead}$ specifies the presence of deadlock. Furthermore, we use the
following shorthand notation (standard):

\medbreak\noindent\hfill$
	\begin{aligned}
		\bot &\enspace\text{instead of}\enspace \neg\top
	\\
		\phi_1 \vee \phi_2 &\enspace\text{instead of}\enspace \neg(\neg\phi_1 \wedge \neg\phi_2)
	\end{aligned}
\qquad
	\begin{aligned}
		\mathsf{AG}(\varphi) &\enspace\text{instead of}\enspace \mathsf{EU}(\top, \neg \varphi)
	\\
		\mathsf{AU}(\varphi_1, \varphi_2) &\enspace\text{instead of}\enspace \neg(\mathsf{EU}(\neg \varphi_2, \neg(\varphi_1 \vee \varphi_2)) \vee \mathsf{EG}(\neg\varphi_2))
	\end{aligned}
$\hfill\strut

\begin{definition}
	Let $\cald \models \varphi$ denote \textit{entailment} of $\varphi$ by $\cald$; it is defined as follows:
	
	\medbreak\noindent\hfill$
		\begin{gathered}[b]
			\dfrac{
			}{
				\cald \models \top
			}
		\qquad
			\dfrac{
				\cald \not\models \varphi
			}{
				\cald \models \neg \varphi
			}
		\qquad
			\dfrac{
				\cald \models \varphi_1
			\qquad
				\cald \models \varphi_2
			}{
				\cald \models \varphi_1 \wedge \varphi_2
			}
		\qquad
			\dfrac{
				\cald \reduxx{}{} \cald'
			\qquad
				\cald \models \varphi
			\qquad
				\cald' \models \mathsf{EG}(\varphi)
			}{
				\cald \models \mathsf{EG}(\varphi)
			}
		\\[\SHIFT]
			\dfrac{
				\cald \models \varphi_2
			}{
				\cald \models \mathsf{EU}(\varphi_1, \varphi_2)
			}
		\qquad
			\dfrac{
				\cald \reduxx{}{} \cald'
			\qquad
				\cald \models \varphi_1
			\qquad
				\cald' \models \mathsf{EU}(\varphi_1, \varphi_2)
			}{
				\cald \models \mathsf{EU}(\varphi_1, \varphi_2)
			}
		\\[\SHIFT]
			\dfrac{
				\cals \reduxxnosmash{\testxx{p}{E}}{\testxx{p}{\datax{true}}} \cals
			}{
				\tupx{\calp, \cals, \calc} \models \testxx{p}{E}
			}
		\qquad
			\dfrac{
				\begin{array}[b]{@{}c@{}}
					\text{for each $\tupx{\calp, \cals, \calc} \reduxx{}{} \tupx{\calp', \cals', \calc'}$}
				\\
					\text{if $\cals(q)(y) \neq \cals'(q)(y)$, then $\tupx{\calp', \cals', \calc'} \models \varphi$}
				\end{array}
			}{
				\tupx{\calp, \cals, \calc} \models \mathsf{AX}_{\atxx{q}{y}}(\varphi)
			}
		\qquad
			\dfrac{
				\smash{\begin{array}[b]{@{}c@{}}
					{\calp \reduxx{}{}}
				\\
					{\tupx{\calp, \cals, \calc} \not\rightarrow}
				\end{array}}
			}{
				\tupx{\calp, \cals, \calc} \models \mathsf{dead}
			}
		\end{gathered}
	$\hfill\llap{$\begin{gathered}\strut\\\strut\qed\end{gathered}$}
\end{definition}

\noindent The rules on the first two lines are the standard ones for CTL. The
first rule on the third line states that a proposition is true if the
corresponding test succeeds. The second rule on the third line states that every
reduction that changes variable $y$ at process $q$ must make $\varphi$ true. The
third rule on the third line states that the presence of deadlock is true if the
set of programs can reduce, but the system cannot (i.e., program reduction is
blocked by stores and/or channels).

In the absence of loops, as in this paper, the model checking problem is
decidable: it is straightforward to adapt classical model checking algorithms
for CTL (e.g., Clarke et al. \cite{DBLP:journals/toplas/ClarkeES86}) to also
support our formulas $\testxx{p}{E}$, $\mathsf{AX}_{\atxx{q}{y}}(\varphi)$, and
$\mathsf{dead}$. If a global program $G$ satisfies operational equivalence, then
it suffices to model check the system that consists of $G$ instead of model
checking the system that consists of $G$'s projections; the former is generally
much more efficient as the state space of $G$'s projections can be exponentially
larger than that of $G$ (due to $\uptau$-reductions of the projections).

\subsection{Examples}

We end this section with some examples. Let:

\medbreak\noindent\hfill$\begin{aligned}
	\cals &= \setx{\rolex{a} \mapsto
	\setx{\datax{hash} \mapsto \datax{0}}, \rolex{b} \mapsto \setx{\datax{x} \mapsto
	\datax{""}, \datax{y} \mapsto \datax{0}}, \rolex{c} \mapsto \setx{\datax{hash}
	\mapsto \datax{0}}}
\\
	\calc &=
	\setxx{\chanxx{p}{q} \mapsto \tupx{\upepsilon, \infty}}{p, q \in
	\setx{\rolex{a}, \rolex{b}, \rolex{c}} \text{ and } p \neq q}
\end{aligned}$\hfill\strut\medbreak

\noindent In words, $\cals$ is an initial family of stores (for Alice, Bob, and
Carol) in which all variables have default values, while $\calc$ is an initial
family of empty channels (between Alice, Bob, and Carol). Furthermore, in addition to
$G_{\rolex{a}\rolex{c}\rolex{b}}^\text{v1}$ in \autoref{sect:intro:problem},
$G_{\rolex{a}\rolex{c}\rolex{b}}^\text{v2}$ in \autoref{sect:intro:contrib}, and
$G_{\rolex{a}\rolex{c}\rolex{b}}^\text{v3}$ in \autoref{sect:lang:props}, let:

\medbreak\noindent\hfill$
	G_{\rolex{a}\rolex{c}\rolex{b}}^\text{v4} = ((\rolex{a} \acq \atxx{\rolex{b}}{\datax{x}}) \async G_{\rolex{a}\rolex{b}}^\text{\autoref{sect:intro:backg}} \async (\rolex{a} \rel \atxx{\rolex{b}}{\datax{x}})) \mer G_{\rolex{c}\rolex{b}}^\text{\autoref{sect:intro:problem}}
\qquad
	G_{\rolex{a}\rolex{c}\rolex{b}}^\text{v5} = (G_{\rolex{a}\rolex{b}}^\text{\autoref{sect:intro:backg}} \async G_{\rolex{c}\rolex{b}}^\text{\autoref{sect:intro:problem}}) \alt (G_{\rolex{c}\rolex{b}}^\text{\autoref{sect:intro:problem}} \async G_{\rolex{a}\rolex{b}}^\text{\autoref{sect:intro:backg}})
$\hfill\strut

\begin{itemize}
	\item Regarding isolation of transactions, the property to be proved can be specified as follows:
	
	\noindent\hfill$
		\varphi = \mathsf{AG}(\mathsf{AX}_{\atrdxx{b}{x}}(\mathsf{AU}(\mathsf{AX}_{\atrdxx{b}{x}}(\bot) \wedge \mathsf{AX}_{\atrdxx{b}{y}}(\bot), \mathsf{AX}_{\atrdxx{b}{y}}(\atrdxx{b}{(md5(x)\,==\,y)}))))
	$\hfill\strut
	
	That is: in all branches, always ($\mathsf{AG}$), if $\datax{x}$ is changed at
	Bob ($\mathsf{AX}_{\atrdxx{b}{x}}$), it is not changed again
	($\mathsf{AX}_{\atrdxx{b}{x}}(\bot)$) until $\datax{y}$ is changed at Bob
	($\mathsf{AX}_{\atrdxx{b}{y}}$) such that $\datax{x}$ and $\datax{y}$ are
	consistent ($\atrdxx{b}{(md5(x)\,==\,y)}$).
	
	System $\tupx{\setx{G_{\rolex{a}\rolex{c}\rolex{b}}^\text{v1}}, \cals, \calc}$
	violates $\varphi$, as informally explained in \autoref{sect:intro:problem}.
	System $\tupx{\setx{G_{\rolex{a}\rolex{c}\rolex{b}}^\text{v2}}, \cals, \calc}$
	satisfies $\varphi$, as Alice and Carol acquire exclusive permission to use
	$\datax{x}$ and $\datax{y}$ at Bob. System
	$\tupx{\setx{G_{\rolex{a}\rolex{c}\rolex{b}}^\text{v3}}, \cals, \calc}$ also
	satisfies $\varphi$: when the system does deadlock, it does so before
	$\datax{x}$ at Bob is changed; when it does not deadlock, Alice and Carol
	acquire exclusive permission. System
	$\tupx{\setx{G_{\rolex{a}\rolex{c}\rolex{b}}^\text{v4}}, \cals, \calc}$
	violates $\varphi$: while
	$G_{\rolex{a}\rolex{b}}^\text{\autoref{sect:intro:backg}}$ runs as an isolated
	transaction (as Alice does acquire exclusive permission),
	$G_{\rolex{c}\rolex{b}}^\text{\autoref{sect:intro:problem}}$ is not (as Carol
	does not). System $\tupx{\setx{G_{\rolex{a}\rolex{c}\rolex{b}}^\text{v5}},
	\cals, \calc}$ satisfies $\varphi$, too, but it violates operational equivalence.
	
	\item Regarding absence of deadlocks, the property to be proved can be
	specified as $\varphi = \mathsf{AG}(\neg \mathsf{dead})$. Systems
	$\tupx{\setx{G_{\rolex{a}\rolex{c}\rolex{b}}^\text{v1}}, \cals, \calc}$,
	$\tupx{\setx{G_{\rolex{a}\rolex{c}\rolex{b}}^\text{v2}}, \cals, \calc}$,
	$\tupx{\setx{G_{\rolex{a}\rolex{c}\rolex{b}}^\text{v4}}, \cals, \calc}$, and
	$\tupx{\setx{G_{\rolex{a}\rolex{c}\rolex{b}}^\text{v5}}, \cals, \calc}$ satisfy
	$\varphi$. In contrast, system $\tupx{\setx{G_{\rolex{a}\rolex{c}\rolex{b}}^\text{v3}},
	\cals, \calc}$ violates $\varphi$.
\end{itemize}

\section{Conclusion}
\label{sect:concl}

\subsection{Related Work}

Advances in choreographic programming were cited in \autoref{sect:intro:backg}.
Outside choreographic programming, closest to our work are mechanisms in the
literature on session types to assure mutual exclusion. In the literature on
\textit{binary} session types, mutual exclusion and related patterns are
supported in the work of Balzer et al. \cite{DBLP:journals/pacmpl/BalzerP17}
(without deadlock freedom) and by Balzer et al. and Kokke et al.
\cite{DBLP:conf/esop/BalzerTP19,DBLP:journals/lmcs/KokkeMW20} (with deadlock
freedom) in the form of typing disciplines for linear and shared channels. In
the literature \textit{multiparty} session types, mutual exclusion is supported
in the work of Voinea et al. \cite{DBLP:conf/ifm/VoineaDG19} in the form of a
typing discipline for linear and shared channels in the special case when
\textit{multiple processes} together implement \textit{a single role}. More
generally, parallel composition has been studied in the context of multiparty
session typing in several ways: through static interleaving of types (e.g.,
\cite{DBLP:conf/popl/HondaYC08,DBLP:journals/jacm/HondaYC16}); through dynamic
interleaving of programs (e.g.,
\cite{DBLP:conf/concur/BettiniCDLDY08,DBLP:journals/mscs/CoppoDYP16}); through a
combination of those two (e.g., in the form of nesting
\cite{DBLP:conf/concur/CarboneLMSW16,DBLP:journals/acta/CarboneMSY17}).

\subsection{This Work}

We presented the design of a new choreographic programming language that
supports isolated transactions among overlapping sets of processes. The first
idea was to track for every variable which processes are permitted to use it.
The second idea was to use model checking to prove isolation. This paper is our
first one in which we pursue these ideas. We believe there is plenty of
room to explore alternative designs and/or refine our work as presented.
Examples include new primitives in the choreographic programming language to
implement programs and new modalities in the temporal logic to specify
properties.

\subsection{Future Work}

On the theoretical side, we see three main avenues. First, we aim to extend the
choreographic programming language with primitives that guarantee isolation and
absence of deadlocks by construction. One possible design is a primitive of the
form ``$\mathbf{isolate}\mspace{4mu}P$'' that implements $P$ as an isolated
transaction. The challenge is to define the operational semantics such that
exclusive permission of variables is automatically acquired as late as possible,
and released as soon as possible, while avoiding deadlocks (e.g., by imposing a
total order on variables). Second, we aim to study an extension of our
choreographic programming language with loops. Third, we aim to investigate
symbolic methods to prove properties.

On the practical side, we are now developing a proof-of-concept implementation
of the design in the form of a compiler from our choreographic programming
language to mCRL2
\cite{DBLP:conf/tacas/CranenGKSVWW13,DBLP:conf/tacas/BunteGKLNVWWW19}. On input
of a global program, the compiler extracts a family of local programs through
projection and translates both the global program and its family to mCRL2
specifications. Using the mCRL2 toolset, we can then check properties of the
global program ($\mu$-calculus versions of our CTL formulas) and operational
equivalence.

\bibliographystyle{eptcs}
\bibliography{bibliography}

\begin{thebibliography}{10}
\providecommand{\bibitemdeclare}[2]{}
\providecommand{\surnamestart}{}
\providecommand{\surnameend}{}
\providecommand{\urlprefix}{Available at }
\providecommand{\url}[1]{\texttt{#1}}
\providecommand{\href}[2]{\texttt{#2}}
\providecommand{\urlalt}[2]{\href{#1}{#2}}
\providecommand{\doi}[1]{doi:\urlalt{https://doi.org/#1}{#1}}
\providecommand{\eprint}[1]{arXiv:\urlalt{https://arxiv.org/abs/#1}{#1}}
\providecommand{\bibinfo}[2]{#2}

\bibitemdeclare{article}{DBLP:journals/pacmpl/BalzerP17}
\bibitem{DBLP:journals/pacmpl/BalzerP17}
\bibinfo{author}{Stephanie \surnamestart Balzer\surnameend} \&
  \bibinfo{author}{Frank \surnamestart Pfenning\surnameend}
  (\bibinfo{year}{2017}): \emph{\bibinfo{title}{Manifest sharing with session
  types}}.
\newblock {\slshape \bibinfo{journal}{Proc. {ACM} Program. Lang.}}
  \bibinfo{volume}{1}(\bibinfo{number}{{ICFP}}), pp.
  \bibinfo{pages}{37:1--37:29}, \doi{10.1145/3110281}.

\bibitemdeclare{inproceedings}{DBLP:conf/esop/BalzerTP19}
\bibitem{DBLP:conf/esop/BalzerTP19}
\bibinfo{author}{Stephanie \surnamestart Balzer\surnameend},
  \bibinfo{author}{Bernardo \surnamestart Toninho\surnameend} \&
  \bibinfo{author}{Frank \surnamestart Pfenning\surnameend}
  (\bibinfo{year}{2019}): \emph{\bibinfo{title}{Manifest Deadlock-Freedom for
  Shared Session Types}}.
\newblock In: {\slshape \bibinfo{booktitle}{{ESOP}}}, {\slshape
  \bibinfo{series}{Lecture Notes in Computer Science}} \bibinfo{volume}{11423},
  \bibinfo{publisher}{Springer}, pp. \bibinfo{pages}{611--639},
  \doi{10.1007/978-3-030-17184-1_22}.

\bibitemdeclare{inproceedings}{DBLP:conf/concur/BettiniCDLDY08}
\bibitem{DBLP:conf/concur/BettiniCDLDY08}
\bibinfo{author}{Lorenzo \surnamestart Bettini\surnameend},
  \bibinfo{author}{Mario \surnamestart Coppo\surnameend},
  \bibinfo{author}{Loris \surnamestart D'Antoni\surnameend},
  \bibinfo{author}{Marco~De \surnamestart Luca\surnameend},
  \bibinfo{author}{Mariangiola \surnamestart Dezani{-}Ciancaglini\surnameend}
  \& \bibinfo{author}{Nobuko \surnamestart Yoshida\surnameend}
  (\bibinfo{year}{2008}): \emph{\bibinfo{title}{Global Progress in Dynamically
  Interleaved Multiparty Sessions}}.
\newblock In: {\slshape \bibinfo{booktitle}{{CONCUR}}}, {\slshape
  \bibinfo{series}{Lecture Notes in Computer Science}} \bibinfo{volume}{5201},
  \bibinfo{publisher}{Springer}, pp. \bibinfo{pages}{418--433},
  \doi{10.1007/978-3-540-85361-9_33}.

\bibitemdeclare{inproceedings}{fm2023}
\bibitem{fm2023}
\bibinfo{author}{Petra \surnamestart van~den Bos\surnameend} \&
  \bibinfo{author}{Sung{-}Shik \surnamestart Jongmans\surnameend}
  (\bibinfo{year}{2023}): \emph{\bibinfo{title}{VeyMont: Parallelising Verified
  Programs Instead of Verifying Parallel Programs}}.
\newblock In: {\slshape \bibinfo{booktitle}{{FM}}}, {\slshape
  \bibinfo{series}{Lecture Notes in Computer Science}} \bibinfo{volume}{14000},
  \bibinfo{publisher}{Springer}, pp. \bibinfo{pages}{321--339},
  \doi{10.1007/978-3-031-27481-7_19}.

\bibitemdeclare{inproceedings}{DBLP:conf/tacas/BunteGKLNVWWW19}
\bibitem{DBLP:conf/tacas/BunteGKLNVWWW19}
\bibinfo{author}{Olav \surnamestart Bunte\surnameend},
  \bibinfo{author}{Jan~Friso \surnamestart Groote\surnameend},
  \bibinfo{author}{Jeroen J.~A. \surnamestart Keiren\surnameend},
  \bibinfo{author}{Maurice \surnamestart Laveaux\surnameend},
  \bibinfo{author}{Thomas \surnamestart Neele\surnameend},
  \bibinfo{author}{Erik~P. \surnamestart de~Vink\surnameend},
  \bibinfo{author}{Wieger \surnamestart Wesselink\surnameend},
  \bibinfo{author}{Anton \surnamestart Wijs\surnameend} \& \bibinfo{author}{Tim
  A.~C. \surnamestart Willemse\surnameend} (\bibinfo{year}{2019}):
  \emph{\bibinfo{title}{The mCRL2 Toolset for Analysing Concurrent Systems -
  Improvements in Expressivity and Usability}}.
\newblock In: {\slshape \bibinfo{booktitle}{{TACAS} {(2)}}}, {\slshape
  \bibinfo{series}{Lecture Notes in Computer Science}} \bibinfo{volume}{11428},
  \bibinfo{publisher}{Springer}, pp. \bibinfo{pages}{21--39},
  \doi{10.1007/978-3-030-17465-1_2}.

\bibitemdeclare{inproceedings}{DBLP:conf/lopstr/CarboneCMM18}
\bibitem{DBLP:conf/lopstr/CarboneCMM18}
\bibinfo{author}{Marco \surnamestart Carbone\surnameend},
  \bibinfo{author}{Lu{\'{\i}}s \surnamestart Cruz{-}Filipe\surnameend},
  \bibinfo{author}{Fabrizio \surnamestart Montesi\surnameend} \&
  \bibinfo{author}{Agata \surnamestart Murawska\surnameend}
  (\bibinfo{year}{2018}): \emph{\bibinfo{title}{Multiparty Classical
  Choreographies}}.
\newblock In: {\slshape \bibinfo{booktitle}{{LOPSTR}}}, {\slshape
  \bibinfo{series}{Lecture Notes in Computer Science}} \bibinfo{volume}{11408},
  \bibinfo{publisher}{Springer}, pp. \bibinfo{pages}{59--76},
  \doi{10.1007/978-3-030-13838-7_4}.

\bibitemdeclare{inproceedings}{DBLP:conf/esop/CarboneHY07}
\bibitem{DBLP:conf/esop/CarboneHY07}
\bibinfo{author}{Marco \surnamestart Carbone\surnameend},
  \bibinfo{author}{Kohei \surnamestart Honda\surnameend} \&
  \bibinfo{author}{Nobuko \surnamestart Yoshida\surnameend}
  (\bibinfo{year}{2007}): \emph{\bibinfo{title}{Structured
  Communication-Centred Programming for Web Services}}.
\newblock In: {\slshape \bibinfo{booktitle}{{ESOP}}}, {\slshape
  \bibinfo{series}{Lecture Notes in Computer Science}} \bibinfo{volume}{4421},
  \bibinfo{publisher}{Springer}, pp. \bibinfo{pages}{2--17},
  \doi{10.1007/978-3-540-71316-6_2}.

\bibitemdeclare{article}{DBLP:journals/toplas/CarboneHY12}
\bibitem{DBLP:journals/toplas/CarboneHY12}
\bibinfo{author}{Marco \surnamestart Carbone\surnameend},
  \bibinfo{author}{Kohei \surnamestart Honda\surnameend} \&
  \bibinfo{author}{Nobuko \surnamestart Yoshida\surnameend}
  (\bibinfo{year}{2012}): \emph{\bibinfo{title}{Structured
  Communication-Centered Programming for Web Services}}.
\newblock {\slshape \bibinfo{journal}{{ACM} Trans. Program. Lang. Syst.}}
  \bibinfo{volume}{34}(\bibinfo{number}{2}), pp. \bibinfo{pages}{8:1--8:78},
  \doi{10.1145/2220365.2220367}.

\bibitemdeclare{inproceedings}{DBLP:conf/concur/CarboneLMSW16}
\bibitem{DBLP:conf/concur/CarboneLMSW16}
\bibinfo{author}{Marco \surnamestart Carbone\surnameend}, \bibinfo{author}{Sam
  \surnamestart Lindley\surnameend}, \bibinfo{author}{Fabrizio \surnamestart
  Montesi\surnameend}, \bibinfo{author}{Carsten \surnamestart
  Sch{\"{u}}rmann\surnameend} \& \bibinfo{author}{Philip \surnamestart
  Wadler\surnameend} (\bibinfo{year}{2016}): \emph{\bibinfo{title}{Coherence
  Generalises Duality: {A} Logical Explanation of Multiparty Session Types}}.
\newblock In: {\slshape \bibinfo{booktitle}{{CONCUR}}}, {\slshape
  \bibinfo{series}{LIPIcs}}~\bibinfo{volume}{59}, \bibinfo{publisher}{Schloss
  Dagstuhl - Leibniz-Zentrum f{\"{u}}r Informatik}, pp.
  \bibinfo{pages}{33:1--33:15}, \doi{10.4230/LIPIcs.CONCUR.2016.33}.

\bibitemdeclare{inproceedings}{DBLP:conf/popl/CarboneM13}
\bibitem{DBLP:conf/popl/CarboneM13}
\bibinfo{author}{Marco \surnamestart Carbone\surnameend} \&
  \bibinfo{author}{Fabrizio \surnamestart Montesi\surnameend}
  (\bibinfo{year}{2013}): \emph{\bibinfo{title}{Deadlock-freedom-by-design:
  multiparty asynchronous global programming}}.
\newblock In: {\slshape \bibinfo{booktitle}{{POPL}}},
  \bibinfo{publisher}{{ACM}}, pp. \bibinfo{pages}{263--274},
  \doi{10.1145/2429069.2429101}.

\bibitemdeclare{article}{DBLP:journals/dc/CarboneMS18}
\bibitem{DBLP:journals/dc/CarboneMS18}
\bibinfo{author}{Marco \surnamestart Carbone\surnameend},
  \bibinfo{author}{Fabrizio \surnamestart Montesi\surnameend} \&
  \bibinfo{author}{Carsten \surnamestart Sch{\"{u}}rmann\surnameend}
  (\bibinfo{year}{2018}): \emph{\bibinfo{title}{Choreographies, logically}}.
\newblock {\slshape \bibinfo{journal}{Distributed Comput.}}
  \bibinfo{volume}{31}(\bibinfo{number}{1}), pp. \bibinfo{pages}{51--67},
  \doi{10.1007/s00446-017-0295-1}.

\bibitemdeclare{article}{DBLP:journals/acta/CarboneMSY17}
\bibitem{DBLP:journals/acta/CarboneMSY17}
\bibinfo{author}{Marco \surnamestart Carbone\surnameend},
  \bibinfo{author}{Fabrizio \surnamestart Montesi\surnameend},
  \bibinfo{author}{Carsten \surnamestart Sch{\"{u}}rmann\surnameend} \&
  \bibinfo{author}{Nobuko \surnamestart Yoshida\surnameend}
  (\bibinfo{year}{2017}): \emph{\bibinfo{title}{Multiparty session types as
  coherence proofs}}.
\newblock {\slshape \bibinfo{journal}{Acta Informatica}}
  \bibinfo{volume}{54}(\bibinfo{number}{3}), pp. \bibinfo{pages}{243--269},
  \doi{10.1007/s00236-016-0285-y}.

\bibitemdeclare{article}{DBLP:journals/toplas/ClarkeES86}
\bibitem{DBLP:journals/toplas/ClarkeES86}
\bibinfo{author}{Edmund~M. \surnamestart Clarke\surnameend},
  \bibinfo{author}{E.~Allen \surnamestart Emerson\surnameend} \&
  \bibinfo{author}{A.~Prasad \surnamestart Sistla\surnameend}
  (\bibinfo{year}{1986}): \emph{\bibinfo{title}{Automatic Verification of
  Finite-State Concurrent Systems Using Temporal Logic Specifications}}.
\newblock {\slshape \bibinfo{journal}{{ACM} Trans. Program. Lang. Syst.}}
  \bibinfo{volume}{8}(\bibinfo{number}{2}), pp. \bibinfo{pages}{244--263},
  \doi{10.1145/5397.5399}.

\bibitemdeclare{article}{DBLP:journals/mscs/CoppoDYP16}
\bibitem{DBLP:journals/mscs/CoppoDYP16}
\bibinfo{author}{Mario \surnamestart Coppo\surnameend},
  \bibinfo{author}{Mariangiola \surnamestart Dezani{-}Ciancaglini\surnameend},
  \bibinfo{author}{Nobuko \surnamestart Yoshida\surnameend} \&
  \bibinfo{author}{Luca \surnamestart Padovani\surnameend}
  (\bibinfo{year}{2016}): \emph{\bibinfo{title}{Global progress for dynamically
  interleaved multiparty sessions}}.
\newblock {\slshape \bibinfo{journal}{Mathematical Structures in Computer
  Science}} \bibinfo{volume}{26}(\bibinfo{number}{2}), pp.
  \bibinfo{pages}{238--302}, \doi{10.1017/S0960129514000188}.

\bibitemdeclare{inproceedings}{DBLP:conf/tacas/CranenGKSVWW13}
\bibitem{DBLP:conf/tacas/CranenGKSVWW13}
\bibinfo{author}{Sjoerd \surnamestart Cranen\surnameend},
  \bibinfo{author}{Jan~Friso \surnamestart Groote\surnameend},
  \bibinfo{author}{Jeroen J.~A. \surnamestart Keiren\surnameend},
  \bibinfo{author}{Frank P.~M. \surnamestart Stappers\surnameend},
  \bibinfo{author}{Erik~P. \surnamestart de~Vink\surnameend},
  \bibinfo{author}{Wieger \surnamestart Wesselink\surnameend} \&
  \bibinfo{author}{Tim A.~C. \surnamestart Willemse\surnameend}
  (\bibinfo{year}{2013}): \emph{\bibinfo{title}{An Overview of the mCRL2
  Toolset and Its Recent Advances}}.
\newblock In: {\slshape \bibinfo{booktitle}{{TACAS}}}, {\slshape
  \bibinfo{series}{Lecture Notes in Computer Science}} \bibinfo{volume}{7795},
  \bibinfo{publisher}{Springer}, pp. \bibinfo{pages}{199--213},
  \doi{10.1007/978-3-642-36742-7_15}.

\bibitemdeclare{inproceedings}{DBLP:conf/ictac/Cruz-FilipeGLMP22}
\bibitem{DBLP:conf/ictac/Cruz-FilipeGLMP22}
\bibinfo{author}{Lu{\'{\i}}s \surnamestart Cruz{-}Filipe\surnameend},
  \bibinfo{author}{Eva \surnamestart Graversen\surnameend},
  \bibinfo{author}{Lovro \surnamestart Lugovic\surnameend},
  \bibinfo{author}{Fabrizio \surnamestart Montesi\surnameend} \&
  \bibinfo{author}{Marco \surnamestart Peressotti\surnameend}
  (\bibinfo{year}{2022}): \emph{\bibinfo{title}{Functional Choreographic
  Programming}}.
\newblock In: {\slshape \bibinfo{booktitle}{{ICTAC}}}, {\slshape
  \bibinfo{series}{Lecture Notes in Computer Science}} \bibinfo{volume}{13572},
  \bibinfo{publisher}{Springer}, pp. \bibinfo{pages}{212--237},
  \doi{10.1007/978-3-031-17715-6_15}.

\bibitemdeclare{inproceedings}{DBLP:conf/fossacs/Cruz-FilipeLM17}
\bibitem{DBLP:conf/fossacs/Cruz-FilipeLM17}
\bibinfo{author}{Lu{\'{\i}}s \surnamestart Cruz{-}Filipe\surnameend},
  \bibinfo{author}{Kim~S. \surnamestart Larsen\surnameend} \&
  \bibinfo{author}{Fabrizio \surnamestart Montesi\surnameend}
  (\bibinfo{year}{2017}): \emph{\bibinfo{title}{The Paths to Choreography
  Extraction}}.
\newblock In: {\slshape \bibinfo{booktitle}{FoSSaCS}}, {\slshape
  \bibinfo{series}{Lecture Notes in Computer Science}} \bibinfo{volume}{10203},
  pp. \bibinfo{pages}{424--440}, \doi{10.1007/978-3-662-54458-7_25}.

\bibitemdeclare{inproceedings}{DBLP:conf/forte/Cruz-FilipeM16}
\bibitem{DBLP:conf/forte/Cruz-FilipeM16}
\bibinfo{author}{Lu{\'{\i}}s \surnamestart Cruz{-}Filipe\surnameend} \&
  \bibinfo{author}{Fabrizio \surnamestart Montesi\surnameend}
  (\bibinfo{year}{2016}): \emph{\bibinfo{title}{Choreographies in Practice}}.
\newblock In: {\slshape \bibinfo{booktitle}{{FORTE}}}, {\slshape
  \bibinfo{series}{Lecture Notes in Computer Science}} \bibinfo{volume}{9688},
  \bibinfo{publisher}{Springer}, pp. \bibinfo{pages}{114--123},
  \doi{10.1007/978-3-319-39570-8_8}.

\bibitemdeclare{inproceedings}{DBLP:conf/sac/Cruz-FilipeM17}
\bibitem{DBLP:conf/sac/Cruz-FilipeM17}
\bibinfo{author}{Lu{\'{\i}}s \surnamestart Cruz{-}Filipe\surnameend} \&
  \bibinfo{author}{Fabrizio \surnamestart Montesi\surnameend}
  (\bibinfo{year}{2017}): \emph{\bibinfo{title}{Encoding asynchrony in
  choreographies}}.
\newblock In: {\slshape \bibinfo{booktitle}{{SAC}}},
  \bibinfo{publisher}{{ACM}}, pp. \bibinfo{pages}{1175--1177},
  \doi{10.1145/3019612.3019901}.

\bibitemdeclare{inproceedings}{DBLP:conf/forte/Cruz-FilipeM17}
\bibitem{DBLP:conf/forte/Cruz-FilipeM17}
\bibinfo{author}{Lu{\'{\i}}s \surnamestart Cruz{-}Filipe\surnameend} \&
  \bibinfo{author}{Fabrizio \surnamestart Montesi\surnameend}
  (\bibinfo{year}{2017}): \emph{\bibinfo{title}{Procedural Choreographic
  Programming}}.
\newblock In: {\slshape \bibinfo{booktitle}{{FORTE}}}, {\slshape
  \bibinfo{series}{Lecture Notes in Computer Science}} \bibinfo{volume}{10321},
  \bibinfo{publisher}{Springer}, pp. \bibinfo{pages}{92--107},
  \doi{10.1007/978-3-319-60225-7_7}.

\bibitemdeclare{article}{DBLP:journals/tcs/Cruz-FilipeM20}
\bibitem{DBLP:journals/tcs/Cruz-FilipeM20}
\bibinfo{author}{Lu{\'{\i}}s \surnamestart Cruz{-}Filipe\surnameend} \&
  \bibinfo{author}{Fabrizio \surnamestart Montesi\surnameend}
  (\bibinfo{year}{2020}): \emph{\bibinfo{title}{A core model for choreographic
  programming}}.
\newblock {\slshape \bibinfo{journal}{Theor. Comput. Sci.}}
  \bibinfo{volume}{802}, pp. \bibinfo{pages}{38--66},
  \doi{10.1016/j.tcs.2019.07.005}.

\bibitemdeclare{inproceedings}{DBLP:conf/sac/Cruz-FilipeMP18}
\bibitem{DBLP:conf/sac/Cruz-FilipeMP18}
\bibinfo{author}{Lu{\'{\i}}s \surnamestart Cruz{-}Filipe\surnameend},
  \bibinfo{author}{Fabrizio \surnamestart Montesi\surnameend} \&
  \bibinfo{author}{Marco \surnamestart Peressotti\surnameend}
  (\bibinfo{year}{2018}): \emph{\bibinfo{title}{Communications in
  choreographies, revisited}}.
\newblock In: {\slshape \bibinfo{booktitle}{{SAC}}},
  \bibinfo{publisher}{{ACM}}, pp. \bibinfo{pages}{1248--1255},
  \doi{10.1145/3167132.3167267}.

\bibitemdeclare{inproceedings}{DBLP:conf/ictac/Cruz-FilipeMP21}
\bibitem{DBLP:conf/ictac/Cruz-FilipeMP21}
\bibinfo{author}{Lu{\'{\i}}s \surnamestart Cruz{-}Filipe\surnameend},
  \bibinfo{author}{Fabrizio \surnamestart Montesi\surnameend} \&
  \bibinfo{author}{Marco \surnamestart Peressotti\surnameend}
  (\bibinfo{year}{2021}): \emph{\bibinfo{title}{Certifying Choreography
  Compilation}}.
\newblock In: {\slshape \bibinfo{booktitle}{{ICTAC}}}, {\slshape
  \bibinfo{series}{Lecture Notes in Computer Science}} \bibinfo{volume}{12819},
  \bibinfo{publisher}{Springer}, pp. \bibinfo{pages}{115--133},
  \doi{10.1007/978-3-030-85315-0_8}.

\bibitemdeclare{inproceedings}{DBLP:conf/itp/Cruz-FilipeMP21}
\bibitem{DBLP:conf/itp/Cruz-FilipeMP21}
\bibinfo{author}{Lu{\'{\i}}s \surnamestart Cruz{-}Filipe\surnameend},
  \bibinfo{author}{Fabrizio \surnamestart Montesi\surnameend} \&
  \bibinfo{author}{Marco \surnamestart Peressotti\surnameend}
  (\bibinfo{year}{2021}): \emph{\bibinfo{title}{Formalising a Turing-Complete
  Choreographic Language in Coq}}.
\newblock In: {\slshape \bibinfo{booktitle}{{ITP}}}, {\slshape
  \bibinfo{series}{LIPIcs}} \bibinfo{volume}{193}, \bibinfo{publisher}{Schloss
  Dagstuhl - Leibniz-Zentrum f{\"{u}}r Informatik}, pp.
  \bibinfo{pages}{15:1--15:18}, \doi{10.4230/LIPIcs.ITP.2021.15}.

\bibitemdeclare{article}{DBLP:journals/scp/EmersonC82}
\bibitem{DBLP:journals/scp/EmersonC82}
\bibinfo{author}{E.~Allen \surnamestart Emerson\surnameend} \&
  \bibinfo{author}{Edmund~M. \surnamestart Clarke\surnameend}
  (\bibinfo{year}{1982}): \emph{\bibinfo{title}{Using Branching Time Temporal
  Logic to Synthesize Synchronization Skeletons}}.
\newblock {\slshape \bibinfo{journal}{Sci. Comput. Program.}}
  \bibinfo{volume}{2}(\bibinfo{number}{3}), pp. \bibinfo{pages}{241--266},
  \doi{10.1016/0167-6423(83)90017-5}.

\bibitemdeclare{inproceedings}{DBLP:conf/forte/GiallorenzoMG18}
\bibitem{DBLP:conf/forte/GiallorenzoMG18}
\bibinfo{author}{Saverio \surnamestart Giallorenzo\surnameend},
  \bibinfo{author}{Fabrizio \surnamestart Montesi\surnameend} \&
  \bibinfo{author}{Maurizio \surnamestart Gabbrielli\surnameend}
  (\bibinfo{year}{2018}): \emph{\bibinfo{title}{Applied Choreographies}}.
\newblock In: {\slshape \bibinfo{booktitle}{{FORTE}}}, {\slshape
  \bibinfo{series}{Lecture Notes in Computer Science}} \bibinfo{volume}{10854},
  \bibinfo{publisher}{Springer}, pp. \bibinfo{pages}{21--40},
  \doi{10.1007/978-3-319-92612-4_2}.

\bibitemdeclare{inproceedings}{DBLP:conf/ecoop/GiallorenzoMPRS21}
\bibitem{DBLP:conf/ecoop/GiallorenzoMPRS21}
\bibinfo{author}{Saverio \surnamestart Giallorenzo\surnameend},
  \bibinfo{author}{Fabrizio \surnamestart Montesi\surnameend},
  \bibinfo{author}{Marco \surnamestart Peressotti\surnameend},
  \bibinfo{author}{David \surnamestart Richter\surnameend},
  \bibinfo{author}{Guido \surnamestart Salvaneschi\surnameend} \&
  \bibinfo{author}{Pascal \surnamestart Weisenburger\surnameend}
  (\bibinfo{year}{2021}): \emph{\bibinfo{title}{Multiparty Languages: The
  Choreographic and Multitier Cases (Pearl)}}.
\newblock In: {\slshape \bibinfo{booktitle}{{ECOOP}}}, {\slshape
  \bibinfo{series}{LIPIcs}} \bibinfo{volume}{194}, \bibinfo{publisher}{Schloss
  Dagstuhl - Leibniz-Zentrum f{\"{u}}r Informatik}, pp.
  \bibinfo{pages}{22:1--22:27}, \doi{10.4230/LIPIcs.ECOOP.2021.22}.

\bibitemdeclare{article}{DBLP:journals/jacm/GlabbeekW96}
\bibitem{DBLP:journals/jacm/GlabbeekW96}
\bibinfo{author}{Rob~J. \surnamestart van Glabbeek\surnameend} \&
  \bibinfo{author}{W.~P. \surnamestart Weijland\surnameend}
  (\bibinfo{year}{1996}): \emph{\bibinfo{title}{Branching Time and Abstraction
  in Bisimulation Semantics}}.
\newblock {\slshape \bibinfo{journal}{J. {ACM}}}
  \bibinfo{volume}{43}(\bibinfo{number}{3}), pp. \bibinfo{pages}{555--600},
  \doi{10.1145/233551.233556}.

\bibitemdeclare{inproceedings}{DBLP:conf/forte/HildebrandtSLDC19}
\bibitem{DBLP:conf/forte/HildebrandtSLDC19}
\bibinfo{author}{Thomas~T. \surnamestart Hildebrandt\surnameend},
  \bibinfo{author}{Tijs \surnamestart Slaats\surnameend},
  \bibinfo{author}{Hugo~A. \surnamestart L{\'{o}}pez\surnameend},
  \bibinfo{author}{S{\o}ren \surnamestart Debois\surnameend} \&
  \bibinfo{author}{Marco \surnamestart Carbone\surnameend}
  (\bibinfo{year}{2019}): \emph{\bibinfo{title}{Declarative Choreographies and
  Liveness}}.
\newblock In: {\slshape \bibinfo{booktitle}{{FORTE}}}, {\slshape
  \bibinfo{series}{Lecture Notes in Computer Science}} \bibinfo{volume}{11535},
  \bibinfo{publisher}{Springer}, pp. \bibinfo{pages}{129--147},
  \doi{10.1007/978-3-030-21759-4_8}.

\bibitemdeclare{article}{DBLP:journals/pacmpl/HirschG22}
\bibitem{DBLP:journals/pacmpl/HirschG22}
\bibinfo{author}{Andrew~K. \surnamestart Hirsch\surnameend} \&
  \bibinfo{author}{Deepak \surnamestart Garg\surnameend}
  (\bibinfo{year}{2022}): \emph{\bibinfo{title}{Pirouette: higher-order typed
  functional choreographies}}.
\newblock {\slshape \bibinfo{journal}{Proc. {ACM} Program. Lang.}}
  \bibinfo{volume}{6}(\bibinfo{number}{{POPL}}), pp. \bibinfo{pages}{1--27},
  \doi{10.1145/3498684}.

\bibitemdeclare{inproceedings}{DBLP:conf/esop/HondaVK98}
\bibitem{DBLP:conf/esop/HondaVK98}
\bibinfo{author}{Kohei \surnamestart Honda\surnameend},
  \bibinfo{author}{Vasco~Thudichum \surnamestart Vasconcelos\surnameend} \&
  \bibinfo{author}{Makoto \surnamestart Kubo\surnameend}
  (\bibinfo{year}{1998}): \emph{\bibinfo{title}{Language Primitives and Type
  Discipline for Structured Communication-Based Programming}}.
\newblock In: {\slshape \bibinfo{booktitle}{{ESOP}}}, {\slshape
  \bibinfo{series}{Lecture Notes in Computer Science}} \bibinfo{volume}{1381},
  \bibinfo{publisher}{Springer}, pp. \bibinfo{pages}{122--138},
  \doi{10.1007/BFb0053567}.

\bibitemdeclare{inproceedings}{DBLP:conf/popl/HondaYC08}
\bibitem{DBLP:conf/popl/HondaYC08}
\bibinfo{author}{Kohei \surnamestart Honda\surnameend}, \bibinfo{author}{Nobuko
  \surnamestart Yoshida\surnameend} \& \bibinfo{author}{Marco \surnamestart
  Carbone\surnameend} (\bibinfo{year}{2008}): \emph{\bibinfo{title}{Multiparty
  asynchronous session types}}.
\newblock In: {\slshape \bibinfo{booktitle}{{POPL}}},
  \bibinfo{publisher}{{ACM}}, pp. \bibinfo{pages}{273--284},
  \doi{10.1145/1328438.1328472}.

\bibitemdeclare{article}{DBLP:journals/jacm/HondaYC16}
\bibitem{DBLP:journals/jacm/HondaYC16}
\bibinfo{author}{Kohei \surnamestart Honda\surnameend}, \bibinfo{author}{Nobuko
  \surnamestart Yoshida\surnameend} \& \bibinfo{author}{Marco \surnamestart
  Carbone\surnameend} (\bibinfo{year}{2016}): \emph{\bibinfo{title}{Multiparty
  Asynchronous Session Types}}.
\newblock {\slshape \bibinfo{journal}{J. {ACM}}}
  \bibinfo{volume}{63}(\bibinfo{number}{1}), pp. \bibinfo{pages}{9:1--9:67},
  \doi{10.1145/2827695}.

\bibitemdeclare{inproceedings}{DBLP:conf/esop/JongmansB22}
\bibitem{DBLP:conf/esop/JongmansB22}
\bibinfo{author}{Sung{-}Shik \surnamestart Jongmans\surnameend} \&
  \bibinfo{author}{Petra \surnamestart van~den Bos\surnameend}
  (\bibinfo{year}{2022}): \emph{\bibinfo{title}{A Predicate Transformer for
  Choreographies - Computing Preconditions in Choreographic Programming}}.
\newblock In: {\slshape \bibinfo{booktitle}{{ESOP}}}, {\slshape
  \bibinfo{series}{Lecture Notes in Computer Science}} \bibinfo{volume}{13240},
  \bibinfo{publisher}{Springer}, pp. \bibinfo{pages}{520--547},
  \doi{10.1007/978-3-030-99336-8_19}.

\bibitemdeclare{inproceedings}{DBLP:conf/lopstr/KjaerCM22}
\bibitem{DBLP:conf/lopstr/KjaerCM22}
\bibinfo{author}{Bj{\o}rn~Angel \surnamestart Kj{\ae}r\surnameend},
  \bibinfo{author}{Lu{\'{\i}}s \surnamestart Cruz{-}Filipe\surnameend} \&
  \bibinfo{author}{Fabrizio \surnamestart Montesi\surnameend}
  (\bibinfo{year}{2022}): \emph{\bibinfo{title}{From Infinity to Choreographies
  - Extraction for Unbounded Systems}}.
\newblock In: {\slshape \bibinfo{booktitle}{{LOPSTR}}}, {\slshape
  \bibinfo{series}{Lecture Notes in Computer Science}} \bibinfo{volume}{13474},
  \bibinfo{publisher}{Springer}, pp. \bibinfo{pages}{103--120},
  \doi{10.1007/978-3-031-16767-6_6}.

\bibitemdeclare{article}{DBLP:journals/lmcs/KokkeMW20}
\bibitem{DBLP:journals/lmcs/KokkeMW20}
\bibinfo{author}{Wen \surnamestart Kokke\surnameend},
  \bibinfo{author}{J.~Garrett \surnamestart Morris\surnameend} \&
  \bibinfo{author}{Philip \surnamestart Wadler\surnameend}
  (\bibinfo{year}{2020}): \emph{\bibinfo{title}{Towards Races in Linear
  Logic}}.
\newblock {\slshape \bibinfo{journal}{Log. Methods Comput. Sci.}}
  \bibinfo{volume}{16}(\bibinfo{number}{4}), \doi{10.23638/LMCS-16(4:15)2020}.

\bibitemdeclare{phdthesis}{M13:phd}
\bibitem{M13:phd}
\bibinfo{author}{Fabrizio \surnamestart Montesi\surnameend}
  (\bibinfo{year}{2013}): \emph{\bibinfo{title}{Choreographic Programming}}.
\newblock \bibinfo{type}{{Ph.{D}. Thesis}}, \bibinfo{school}{IT University of
  Copenhagen}.
\newblock
  \bibinfo{note}{\href{https://www.fabriziomontesi.com/files/choreographic-programming.pdf}{https://www.fabriziomontesi.com/files/choreographic-programming.pdf}}.

\bibitemdeclare{inproceedings}{DBLP:conf/concur/MontesiY13}
\bibitem{DBLP:conf/concur/MontesiY13}
\bibinfo{author}{Fabrizio \surnamestart Montesi\surnameend} \&
  \bibinfo{author}{Nobuko \surnamestart Yoshida\surnameend}
  (\bibinfo{year}{2013}): \emph{\bibinfo{title}{Compositional Choreographies}}.
\newblock In: {\slshape \bibinfo{booktitle}{{CONCUR}}}, {\slshape
  \bibinfo{series}{Lecture Notes in Computer Science}} \bibinfo{volume}{8052},
  \bibinfo{publisher}{Springer}, pp. \bibinfo{pages}{425--439},
  \doi{10.1007/978-3-642-40184-8_30}.

\bibitemdeclare{inproceedings}{DBLP:conf/coordination/PredaGGLM15}
\bibitem{DBLP:conf/coordination/PredaGGLM15}
\bibinfo{author}{Mila~Dalla \surnamestart Preda\surnameend},
  \bibinfo{author}{Maurizio \surnamestart Gabbrielli\surnameend},
  \bibinfo{author}{Saverio \surnamestart Giallorenzo\surnameend},
  \bibinfo{author}{Ivan \surnamestart Lanese\surnameend} \&
  \bibinfo{author}{Jacopo \surnamestart Mauro\surnameend}
  (\bibinfo{year}{2015}): \emph{\bibinfo{title}{Dynamic Choreographies - Safe
  Runtime Updates of Distributed Applications}}.
\newblock In: {\slshape \bibinfo{booktitle}{{COORDINATION}}}, {\slshape
  \bibinfo{series}{Lecture Notes in Computer Science}} \bibinfo{volume}{9037},
  \bibinfo{publisher}{Springer}, pp. \bibinfo{pages}{67--82},
  \doi{10.1007/978-3-319-19282-6_5}.

\bibitemdeclare{article}{DBLP:journals/corr/PredaGGLM16}
\bibitem{DBLP:journals/corr/PredaGGLM16}
\bibinfo{author}{Mila~Dalla \surnamestart Preda\surnameend},
  \bibinfo{author}{Maurizio \surnamestart Gabbrielli\surnameend},
  \bibinfo{author}{Saverio \surnamestart Giallorenzo\surnameend},
  \bibinfo{author}{Ivan \surnamestart Lanese\surnameend} \&
  \bibinfo{author}{Jacopo \surnamestart Mauro\surnameend}
  (\bibinfo{year}{2017}): \emph{\bibinfo{title}{Dynamic Choreographies: Theory
  And Implementation}}.
\newblock {\slshape \bibinfo{journal}{Log. Methods Comput. Sci.}}
  \bibinfo{volume}{13}(\bibinfo{number}{2}), \doi{10.1007/BF01221097}.

\bibitemdeclare{inproceedings}{DBLP:conf/sle/PredaGLMG14}
\bibitem{DBLP:conf/sle/PredaGLMG14}
\bibinfo{author}{Mila~Dalla \surnamestart Preda\surnameend},
  \bibinfo{author}{Saverio \surnamestart Giallorenzo\surnameend},
  \bibinfo{author}{Ivan \surnamestart Lanese\surnameend},
  \bibinfo{author}{Jacopo \surnamestart Mauro\surnameend} \&
  \bibinfo{author}{Maurizio \surnamestart Gabbrielli\surnameend}
  (\bibinfo{year}{2014}): \emph{\bibinfo{title}{{AIOCJ:} {A} Choreographic
  Framework for Safe Adaptive Distributed Applications}}.
\newblock In: {\slshape \bibinfo{booktitle}{{SLE}}}, {\slshape
  \bibinfo{series}{Lecture Notes in Computer Science}} \bibinfo{volume}{8706},
  \bibinfo{publisher}{Springer}, pp. \bibinfo{pages}{161--170},
  \doi{10.1007/978-3-319-11245-9_9}.

\bibitemdeclare{inproceedings}{DBLP:conf/ifm/VoineaDG19}
\bibitem{DBLP:conf/ifm/VoineaDG19}
\bibinfo{author}{A.~Laura \surnamestart Voinea\surnameend},
  \bibinfo{author}{Ornela \surnamestart Dardha\surnameend} \&
  \bibinfo{author}{Simon~J. \surnamestart Gay\surnameend}
  (\bibinfo{year}{2019}): \emph{\bibinfo{title}{Resource Sharing via
  Capability-Based Multiparty Session Types}}.
\newblock In: {\slshape \bibinfo{booktitle}{{IFM}}}, {\slshape
  \bibinfo{series}{Lecture Notes in Computer Science}} \bibinfo{volume}{11918},
  \bibinfo{publisher}{Springer}, pp. \bibinfo{pages}{437--455},
  \doi{10.1007/978-3-030-34968-4_24}.

\end{thebibliography}
\end{document}